\documentclass[
 reprint,amsmath,amssymb,aps,prb,eps,superscriptaddress,twocolumn,10pt
]{revtex4-2}

\usepackage{graphicx}% Include figure files
\usepackage{dcolumn}% Align table columns on decimal point
\usepackage{grffile}% Extended file name support for graphics (dots)
\usepackage{bm}% bold math
\usepackage[colorlinks=true,
            linkcolor=blue,
            urlcolor=blue,
            citecolor=blue]{hyperref}
\usepackage{xcolor}
\usepackage{placeins}
\usepackage{changes}
\usepackage{enumitem}
\usepackage{blindtext}
\usepackage{notes2bib}

\definechangesauthor[name=Markus Richter, color=teal]{MR}
\definechangesauthor[name=Johannes Graspeuntner, color=blue]{JG}

\begin{document}

\preprint{APS/123-QED}

\title{Comparing the effective enhancement of local and non-local spin-orbit couplings on honeycomb lattices due to strong electronic correlations}

\author{Markus Richter}
\affiliation{Institute of Theoretical and Computational Physics, Graz University of Technology, NAWI Graz, Petersga{\ss}e 16, Graz, 8010, Austria.}
\author{Johannes Graspeuntner}
\affiliation{Institute of Theoretical and Computational Physics, Graz University of Technology, NAWI Graz, Petersga{\ss}e 16, Graz, 8010, Austria.}
\author{Thomas Sch\"afer}
\affiliation{Max-Planck-Institut f\"ur Festk\"orperforschung, Heisenbergstra{\ss}e 1, 70569 Stuttgart, Germany}
\author{Nils Wentzell}
\affiliation{Center for Computational Quantum Physics,  Flatiron Institute,  162 5th Avenue,  New York,  NY 10010,  USA}
\author{Markus Aichhorn}
\email{aichhorn@tugraz.at}
\affiliation{Institute of Theoretical and Computational Physics, Graz University of Technology, NAWI Graz, Petersga{\ss}e 16, Graz, 8010, Austria.}

\newcommand{\appropto}{\mathrel{\vcenter{
  \offinterlineskip\halign{\hfil$##$\cr
    \propto\cr\noalign{\kern2pt}\sim\cr\noalign{\kern-2pt}}}}}

\date{\today}

\begin{abstract}
We investigate the interplay of electronic correlations and spin-orbit coupling (SOC) for a one-band and a two-band honeycomb lattice model. The main difference between the two models concerning SOC is that in the one-band case the SOC is a purely non-local term in the basis of the $p_z$ orbitals, whereas in the two-band case with $p_x$ and $p_y$ as basis functions it is purely local. In order to grasp the correlation effects on non-local spin-orbit coupling, we apply the TRILEX approach that allows to calculate non-local contributions to the self-energy approximately. For the two-band case we apply dynamical mean-field theory. 
In agreement with previous studies, we find that for all parameter values in our study, the effect of correlations on the spin-orbit coupling strength is that the bare effective SOC parameter is increased. However, this increase is much weaker in the non-local than in the local SOC case. 
Concerning the TRILEX method, we introduce the necessary formulas for calculations with broken SU(2) symmetry.
\end{abstract}

\maketitle

\section{Introduction}

The interplay of spin-orbit coupling (SOC) and strong electronic correlations has attracted a lot of attention in recent years~\cite{krempa_review}. 

On the one hand, mainly triggered by the discovery of insulating behavior in Sr$_2$IrO$_4$~\cite{moon_2008}, a lot of research has been devoted to understand, how SOC can influence the degree of correlations in materials with heavy transition metal ions~\cite{martins_2011,zhang_2013,guoren_2016,martins_2017,triebl_spin-orbit_2018}. In these iridate compounds, for instance, SOC has been identified as the driving force to reach the strongly correlated insulating state~\cite{martins_2011,arita_2012}. Without SOC, the relevant electronic orbitals would be the t$_{2g}$ orbitals. The SOC now splits this manifold into $J_{\textrm eff}=3/2$ and $J_{\textrm eff}=1/2$ orbitals, where the latter ones become half-filled, leading to a strongly-correlated spin-orbital polarized electronic state~\cite{martins_2017}. 

On the other hand, one can also ask the inverse question, which is to what extent the spin-orbit coupling is changed by the inclusion of electronic correlations. This is also important in the context of correlated topological insulators~\cite{krempa_review}. In topological insulators in general, SOC is the main driving force for topological properties, and different topological phases can occur as function of the spin-orbit coupling parameter $\lambda_{SO}$, interaction strength $U$, and/or other parameters like crystal or external field effects~\cite{triebl_topological_2016}. If parameters are such that the system is close to a phase transition, external perturbations might be able to even switch between topological and non-topological states, as shown, e.g., in Ref.~\cite{wehling2021}. 

Quite generally, it has been found both in model as well as material-related studies~\cite{liu_coulomb-enhanced_soc_2008, behrmann_multiorbital_2012, zhang_fermi_2016, bunemann_interplay_2016, kim_spin-orbit_2018, triebl_spin-orbit_2018, linden_imaginary-time_2020} that electronic correlations \textit{enhance} the SOC strength, leading to an effective SOC $\lambda_{\text{eff}}>\lambda_{SO}$.  For ruthenates, this has also been confirmed experimentally~\cite{tamai_ARPES_2019}. We want to note that the concept of effective single-particle parameters is not at all limited to the case of SOC presented here. For instance, band-width (hopping) renormalization due to non-local interactions~\cite{Ayral_Fock_diagram_2017,in_t_Veld_2019} has been studied.

In this work, we investigate the effective SOC for two similar, but yet complementary systems. First, we want to study  
this behavior for a model, where the spin-orbit coupling is a \textit{non-local term} in the Hamiltonian. We choose the Kane-Mele-Hubbard (KMH) model \citep{kane_mele_quantum_spin_hall_2005,kane_mele_z2_2005,rachel_topological_2010}  as an example. It has been a popular model to investigate on the effects of correlations to a topological phase
\citep{hohenadler_correlation_qsh_2011, lee_interaction-effects_qsh_2011, hohenadler_quantum_KMH_2012, assaad_topological_2013, hohenadler_correlation_2dtopological_2013, hung_interaction_2014, lai_effects_KMH_2014, yu_mott_topological_phasetransition_2011, budich_fluctuation-induced_qsh_qah_2012, laubach_rashba_2014, miyakoshi_coexistence_2015, wu_qsh_2012, grandi_topological_cpt_2015, chen_cellular_2015, parisen_toldin_fermionic_2015, rachel_quantum_2016, triebl_topological_2016, quan_phase_2017, li_edge_2017, mishra_magnetic_2018, jiang_antiferromagnetic_2018, du_phase_2018, mishra_topological_2018, novelli_failure_2019, losada_ultrafast_2019, du_competition_2020}, but has never been studied concerning the enhancement of spin-orbit coupling due to correlations. 

Second, we construct a two-band honeycomb model motivated by bismuthene \cite{science_paper_Bismuthene_on_SiC,freitas_topological_2015}, 
which can be viewed as the two-orbital variant of the graphene honeycomb lattice, but where the SOC term is again local. Comparing these two cases with similar structure and both at half-filling, we can quantify how non-local spin-orbit couplings respond to electronic correlations as opposed to the case of spin-orbit couplings. As we will explain in more detail below, this enhancing effect is calculated from the zero-energy limit of the self-energy.

While in order to investigate the enhancement of a local spin-orbit coupling dynamical mean-field theory (DMFT)~\cite{georges_dmft_1996} works very well and is, hence, our method of choice for bismuthene, this is no longer sufficient for the KMH model. Due to the local approximation of the self-energy in DMFT, its contribution to the non-local SOC is a-priori zero. In order to get a non-zero contribution we have to go beyond the local approximation of DMFT. Therefore, we adopt the TRILEX approach developed by Ayral and Parcollet~\cite{ayral_trilex_2015, ayral_trilex_2016}, which uses a local approximation of the electron-boson vertex and, hence, recovers a non-local contribution to the self-energy. 
Since the spin-orbit coupling term in the Kane-Mele Hubbard model breaks SU(2)-symmetry, we need to develop the necessary TRILEX formulas for this case, which is possible within the original TRILEX formulation. However, the SU(2)-broken version of this original TRILEX approach can violate hermiticity for channel-off-diagonal components. As detailed in Sec \ref{sec:trilex} we therefore employ a variant of the method coined TRILEX $\Lambda^2$~\citep{double_Lambda4} which resolves this issue.

The generalization of the TRILEX formulas to SU(2)-broken is not only necessary in the present case (where we do paramagnetic calculations with SOC) but also for other calculations inside a spin-symmetry broken phase, such as an antiferromagnet.

\section{Models}

\begin{figure}
	\begin{center}
		\includegraphics[width=0.6\columnwidth]{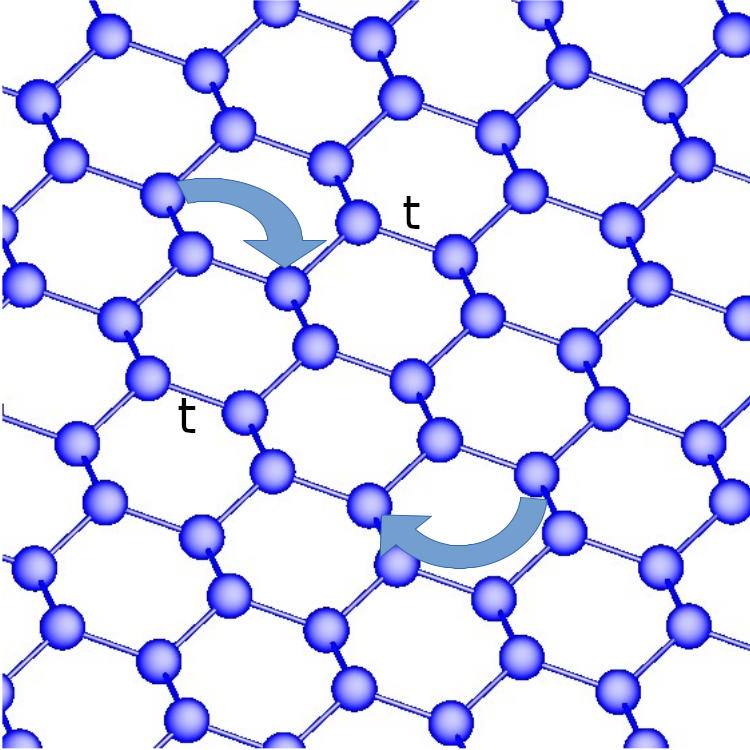}
		\vspace*{0.5cm}
		\includegraphics[width=0.6\columnwidth]{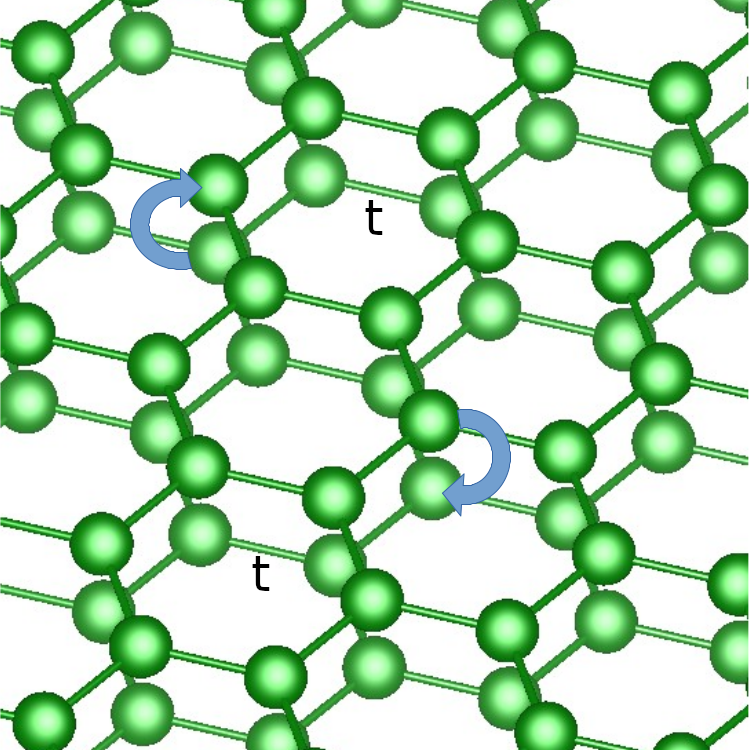}
		\caption{Sketch of the lattice structure of the two honeycomb lattices. Top: Single-band honeycomb lattice for the KMH model with non-local SOC. Bottom: Two-orbital lattice, relevant for bismuthene with a local SOC term. The bonds with strongest hopping term in the models are indicated by $t$, and the SOC is also indicated by arrows.}
		\label{fig:kmh_bis_lattice}
	\end{center}
\end{figure}

\subsection{One-band model: KMH}

\begin{figure}
	\begin{center}
	\includegraphics[width=0.49\columnwidth]{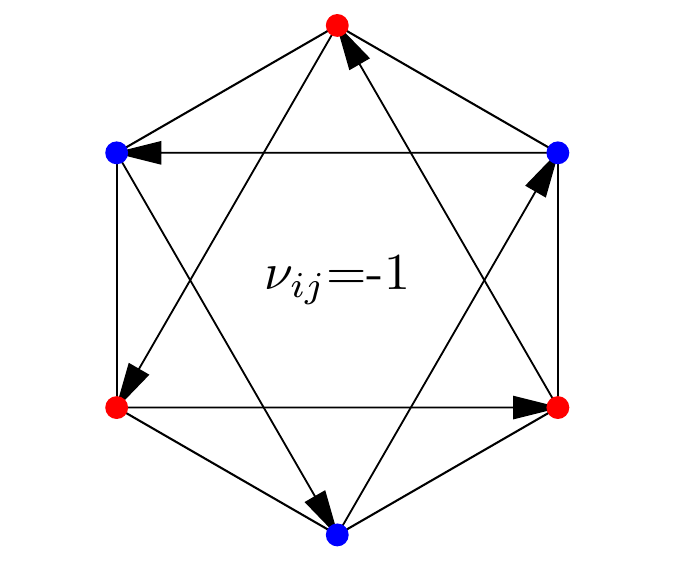}
	\includegraphics[width=0.49\columnwidth]{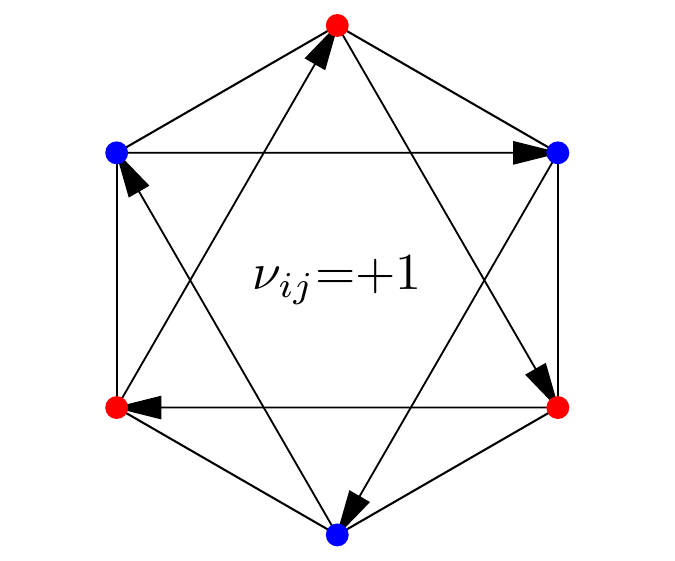}
	\caption{Illustration of the sign $\nu_{ij}$ for the Kane-Mele-Model. Every right turn gives $\nu_{ij}$ the value $+1$, and every left turn the value $-1$.}
	\label{fig:kmh_complex_hopping}
	\end{center}
\end{figure}

The KMH model describes locally interacting electrons on the two-dimensional honeycomb lattice with nearest-neighbor hopping and spin-orbit coupling. The Hamiltonian reads
\begin{equation}
\label{eq:kmh_Hamiltonian}
\begin{aligned}
H = &- t\sum_{\left<ij\right>,\sigma} c_{i\sigma}^{\dagger}c^{\phantom{\dagger}}_{j\sigma} + U \sum_{i}n_{i\uparrow}n_{i\downarrow} \\&+ 
i\lambda_{SO}^{1B}\sum_{\left<\left<ij\right>\right>,\sigma,\sigma'}\nu_{ij}c_{i\sigma}^{\dagger}\sigma^{z}_{\sigma\sigma'}c^{\phantom{\dagger}}_{j\sigma'} ,
\end{aligned}
\end{equation}
with the creation (annihilation) operator $c^{\dagger}_{i\sigma}$ ($c^{\phantom{\dagger}}_{i\sigma}$) creating (annihilating) an electron with spin $\sigma$ at site $i$. The density operator is $n_{i\sigma} = c^{\dagger}_{i\sigma}c^{\phantom{\dagger}}_{i\sigma}$, $\sigma^z$ is the Pauli matrix, and $\nu_{ij}$ gives $+1$ for every right turn and $-1$ for every left turn as illustrated in Fig.~\ref{fig:kmh_complex_hopping}. The first term is a simple tight binding term, including just next-nearest neighbors. The second term is the well-known (purely local) Hubbard interaction. The last term includes just next-nearest neighbors and relates to the SOC in graphene. Here SOC is a purely non-local quantity, and its strength is given by $\lambda_{SO}^{1B}$, where we use the superscript $1B$ to refer to the 1-band model.

The non-interacting case ($U=0$) can be solved exactly by a Fourier transformation of the Hamiltonian. Therefore we are fixing the unit vectors of the lattice to $\mathbf{a_1} = \left(3/2,\sqrt{3}/2\right)$ and $\mathbf{a_2} = \left(0,\sqrt{3}\right)$.
The honeycomb lattice has a natural bipartite structure, i.e., it consists of two interpenetrating triangular sub-lattices A and B (see also Fig.~\ref{fig:kmh_complex_hopping}). As the SOC term breaks SU(2) symmetry, a full basis-set for $H(\textbf{k})$ consists of four different species with a defined spin-direction. Taking $\Psi^{\dagger (S_z)}_{\mathbf{k}} = \left(a_{\mathbf{k},\uparrow}^{\dagger},b_{\mathbf{k},\uparrow}^{\dagger},a_{\mathbf{k},\downarrow}^{\dagger},b_{\mathbf{k},\downarrow}^{\dagger}\right)$ as a basis-set with $a_{\mathbf{k},\sigma}^{\dagger}$ living on sub-lattice A and $b_{\mathbf{k},\sigma}^{\dagger}$ on sub-lattice B, the Hamilton matrix reads 
\begin{equation}
\label{eq:kmh_H_0_k}
H_0^{(Sz)}(\mathbf{k}) = \begin{pmatrix}
\gamma_{\mathbf{k}} & -g_{\mathbf{k}}^{\phantom{*}} & 0 & 0  \\
-g_{\mathbf{k}}^* & -\gamma_{\mathbf{k}} & 0 & 0 \\
0 & 0 & -\gamma_{\mathbf{k}}& -g^{\phantom{*}}_{\mathbf{k}}\\
0 & 0 & -g_{\mathbf{k}}^* & \gamma^{\phantom{*}}_{\mathbf{k}}
\end{pmatrix},
\end{equation}
with
\begin{equation}
\label{eq:kmh_g_k}
g^{\phantom{*}}_{\mathbf{k}} = t e^{i\frac{\sqrt{3}}{2}k_y}\left[e^{i \frac{3}{2}k_x} + 2 \cos\left(\frac{\sqrt{3} k_y}{2}\right)\right]
\end{equation}
and
\begin{equation}
\label{eq:kmh_gamma_k}
\gamma_{\mathbf{k}} = 2\lambda^{1B}_{SO}\left[2\sin\left(\frac{\sqrt{3}}{2}k_y\right)\cos\left(\frac{3}{2}k_x\right)-\sin\left(\sqrt{3}k_y\right)\right].
\end{equation}
The two inequivalent high-symmetry points lie at $\mathbf{K} = \left(2\pi/3, 2\pi/(3\sqrt{3})\right)$ and $\mathbf{K'} = \left(2\pi/3, -2\pi/(3\sqrt{3})\right)$. At these high-symmetry points, in the absence of $\lambda^{1B}_{SO}$, we find Dirac cones. Finite spin-orbit coupling opens up a gap of size $6\sqrt{3}\lambda^{1B}_{SO}$, resulting in a topological phase. A more detailed discussion of the KMH Model can be found, e.g., in \citep{rachel_topological_2010, hohenadler_correlation_2dtopological_2013}.

\subsection{Two-band model: Bismuthene}

Bismuth has five valence electrons and, therefore, forms a buckled hexagonal lattice with all the $p$-orbitals being near the Fermi edge. When passivating this sheet of bismuth atoms with either a suitable substrate or hydrogen, not only does the buckling vanish and the system becomes planar, but the $p_z$-orbitals are also shifted away from the Fermi energy. This happens because the excess valence electron binds with the passivating molecules, resulting in Dirac cones at the K-points for the $p_x$- and the $p_y$-orbitals. As a result, the system has to be described as an effective two-band model. The theoretical background and experimental verifications of substrate-passivated bismuthene have been presented by Reis~\textit{et al.} \citep{science_paper_Bismuthene_on_SiC}.  In this work, the passivation is performed by adding hydrogen atoms to the bismuthene sheet, as was done by Freitas~\textit{et al.} \citep{freitas_topological_2015}. 
Due to the high atomic number of bismuth, spin-orbit coupling is large and opens a considerable gap at the Fermi level. Note that performing the passivation with hydrogen atoms instead of using a SiC substrate results in a direct gap instead of an indirect one.

As already mentioned, the important effect of the passivation of bismuthene is that it removes the $p_z$ orbital from the Fermi level, and the relevant bands are instead coming from $p_x$ and $p_y$ orbitals. In this two-orbital basis, it has been shown that the SOC term has local matrix elements and can be written as a local operator \citep{science_paper_Bismuthene_on_SiC},
\begin{equation}
\label{eq:bis_H_SOC}
H_{SO} = \frac{1}{2} \begin{pmatrix}
0 & i & 0 & 0 \\
-i & 0 & 0 & 0 \\
0 & 0 & 0 & -i \\
0 & 0 & i & 0 
\end{pmatrix}.
\end{equation}
The non-interacting Hamiltonian for our two-band model is constructed as follows. We perform density-functional theory calculations for bismuthene without SOC included using the VASP code \citep{VASP1,VASP2,VASP3,VASP4,VASP5}. To obtain the atomic positions as input data, we used the knowledge that bismuthene has a hexagonal structure just like graphene and took the atomic distances from Freitas~\textit{et al.} \citep{freitas_topological_2015}. We added a hydrogen atom to each of the bismuth atoms, for sublattice A in the positive and for sublattice B in the negative $z$-direction. 
We used a $\Gamma$-centered 27$\times$27$\times$1 $\mathbf{k}$-point grid and a basis-set cutoff ENCUT = $200$\,eV. 
We then construct a non-interacting Hamiltonian $H_0^\text{noSOC}(\mathbf{k})$ by a maximally-localized Wannier90 projection~\cite{Wannier90} to the $p_x$ and $p_y$ orbitals. We provide the output of this projection~\citep{hr_bismuthene} in a machine-readable format in the supplementary material of this publication. Other formats of the data are available from the authors upon request. The full non-interacting Hamiltonian is then given by  
\begin{equation}
\label{eq:bis_H_0}
H_{0}(\mathbf{k}) = H_{0}^\text{noSOC}(\mathbf{k}) + \lambda_{SO}^{2B}  H_{SO},
\end{equation}
where we use $\lambda_{SO}^{2B}$ as SOC strength and the superscript $2B$ to refer to the two-band model. 

In order to check whether a purely local SOC term is a good approximation, we also did DFT calculations including the SOC directly. We find that by choosing $\lambda_{SO}^{2B}=0.435$\,eV in Eq.~\eqref{eq:bis_H_0} the agreement between the band structures obtained in the two ways is excellent. Approximately the same value for the SOC was derived in the work of Reis~\textit{et al.} \citep{science_paper_Bismuthene_on_SiC}. The full non-interacting Hamiltonian is then used as input for the DMFT calculations. As interaction part of the Hamiltonian we use a Kanamori-Hamiltonian of the form 
\begin{equation}
\label{eq:bis_H_int}
\begin{aligned}
H_{int} &= U \sum_{m} n_{m \uparrow} n_{m \downarrow} +  U' \sum_{m \neq m'} n_{m \uparrow} n_{m' \downarrow} \\ 
&+ (U'- J) \sum_{m<m',\sigma} n_{m\sigma}n_{m' \sigma} \\
& + J \sum_{m \neq m'}c_{m \uparrow}^\dagger c_{m' \downarrow}^\dagger c_{m \downarrow} c_{m' \uparrow} \\
&+ J \sum_{m \neq m'} c_{m \uparrow}^\dagger c_{m \downarrow}^\dagger c_{m' \downarrow} c_{m' \uparrow},
\end{aligned}
\end{equation}
where $U$ is the Coulomb interaction and $J$ denotes the Hund's coupling, and $U^{\prime}$ is defined as $U - 2J$ due to rotational symmetry. For the calculations a we choose a fixed ratio of $J = 0.2 U$.

The DMFT calculation has been performed with the TRIQS library \cite{triqs} and the continuous-time quantum Monte Carlo impurity solver as implemented in the TRIQS/CTHYB application~\cite{cthyb_triqs}. 

\subsection{Making the models comparable}

\begin{figure}
	\begin{center}
		\includegraphics[width=\columnwidth]{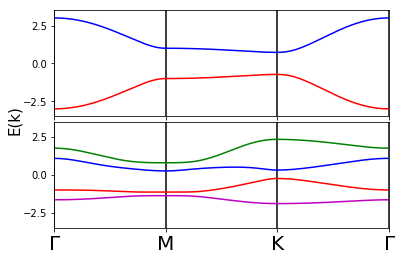}
		\caption{Top: Dispersion relation for the non-interacting Kane-Mele Model with $\lambda_{SO}^{1B} = 0.14t$. Bottom: Dispersion relation for the non-interacting two-band model (bismuthene) with $\lambda_{SO}^{2B} = 0.21t$. For both band structures, we set the nearest-neighbor orbital-diagonal hopping to $t=1$, in order to set a common unit of energy.}
		\label{fig:Dispersion_comp}
	\end{center}
\end{figure}

Since we are calculating the non-interacting band structure for the two-band model for bismuthene from DFT, we first normalize this band structure such that the largest matrix element in the tight-binding Hamiltonian is $t=1$. This largest element is the nearest-neighbor orbital-diagonal hopping, as shown in Fig.~\ref{fig:kmh_bis_lattice}. This allows us to use this matrix element $t=1$ as unit of energy throughout the paper.

Next, we have to set the SOC strengths in the two models. In both cases, we choose this strength such that the ratio of the gap $\Delta$ opened by SOC to the relevant band width $W$ is $\Delta/W = 1/4$. Note that for the two-band model we use the band-width of the two innermost bands for this comparison. This results in $\lambda_{SO}^{1B}=0.14t$ and $\lambda_{SO}^{2B}=0.21t$. Note that this value for the SOC in the two-band case is very close to the actual value determined from DFT for Bismuthene, which in the same units reads $\lambda_{SO}^{DFT}=0.24t$. The dispersion relations for the non-interacting models are shown in Fig.~\ref{fig:Dispersion_comp}.

\section{Self-Energy contribution to the Effective SOC}

\subsection{One-band model and non-local SOC}
\label{subsec:one_band_model_and_non_local_soc}

The self-energy of the KMH-model (Eq.~\eqref{eq:kmh_Hamiltonian}) in the paramagnetic phase has the following form:
\begin{equation}
\label{eq:kmh_sigma_sz}
\setlength{\arraycolsep}{0.02pt}
\Sigma^{(Sz)} = \begin{pmatrix}
\Sigma^{d}\hspace{-0.1cm}+\hspace{-0.1cm}\Sigma^{SO} & \Sigma_{AB} & 0 & 0  \\
\Sigma_{BA} & \Sigma^{d}\hspace{-0.1cm}-\hspace{-0.1cm}\Sigma^{SO} & 0 & 0 \\
0 & 0 & \Sigma^{d}\hspace{-0.1cm}-\hspace{-0.1cm}\Sigma^{SO}& \Sigma_{AB}\\
0 & 0 & \Sigma_{BA} & \Sigma^{d}\hspace{-0.1cm}+\hspace{-0.1cm}\Sigma^{SO}
\end{pmatrix}
\end{equation}
Here, $\mathbf{k}$ and $i\omega_n$ indices have been dropped to improve the readability. On the diagonal we find an ordinary diagonal part $\Sigma^d$ and another part $\Sigma^{SO}$ that breaks SU(2) symmetry. 

The motivation to use this nomenclature is the following. For a model without SOC, the contribution $\Sigma^{SO}$ is obviously zero and one is left only with $\Sigma^d$. When SOC is included, we additionally get a contribution that adopts the symmetry of $\gamma_\mathbf{k}$ (Eq.~\eqref{eq:kmh_gamma_k}) in the Hamiltonian~\eqref{eq:kmh_H_0_k}. The symmetry of these terms in the Brillouin zone matches $\gamma_{\mathbf{k}} = -\gamma_{-\mathbf{k}}$, and in spin- and sublattice space we obtain a sign structure corresponding to the 'diag($\gamma_\mathbf{k}$,-$\gamma_\mathbf{k}$,-$\gamma_\mathbf{k}$,$\gamma_\mathbf{k}$)' part of Hamiltonian~\eqref{eq:kmh_H_0_k}.
Furthermore, this structure also implies that the local contribution of $\Sigma^{SO}$, i.e. summed over all momenta, is zero, and does not produce any spin splitting on the impurity/atom.

All TRILEX self-consistent calculations are done using the block structure of Eq.~\eqref{eq:kmh_sigma_sz}. However, in the postprocessing we need to disentangle $\Sigma^{SO}$ from $\Sigma^{d}$ in order to define an effective SOC. To do so, we can rotate the spin-subspace from the vertical (z) into the horizontal axis (x/y) using the transformation matrix
\begin{equation}
\label{eq:kmh_trans_matrix_sz_to_sx}
T = e^{-i\pi/4\sigma_y}\times \begin{pmatrix}1&0\\0&1\end{pmatrix}
\end{equation}
This leads the following matrix form of the self-energy
\begin{equation}
\label{eq:kmh_sigma_sx}
\Sigma^{(Sx)} = T\Sigma^{(Sz)}T^{-1} = \begin{pmatrix}
\Sigma^{d}  & \Sigma_{AB} & \Sigma^{SO} & 0  \\
\Sigma_{BA} & \Sigma^{d}  & 0 & -\Sigma^{SO}\\
\Sigma^{SO} & 0 & \Sigma^{d} & \Sigma_{AB}\\
0 & -\Sigma^{SO} & \Sigma_{BA} & \Sigma^{d}
\end{pmatrix}
\end{equation}
where the contribution due to SOC is now well separated on the off-diagonal.

From this self-energy $\Sigma^{SO}$ we can define an effective spin-orbit coupling. Some care has to be taken because of the sign factor $\nu_{ij}$ in the Hamiltonian, which gives a factor $+1$ or $-1$ depending on the direction of the bond, see Fig.~\ref{fig:kmh_complex_hopping}.  We can 
define the static enhancement of the SOC as 
\begin{equation}
\label{eq:kmh_lambda_eff}
\lambda_{\text{eff}}^{1B} = \lambda_{SO}^{1B} + \nu_{ij}\,\text{Im}\,\Sigma^{SO}(\textbf{r}-\textbf{r}' =\mathbf{a}_{ij},i\omega = i\omega_0)
\end{equation}
with $\mathbf{a}_{ij}$ being the vector of the honeycomb lattice along a next-nearest neighbor bond, where $\nu_{ij}=\pm 1$. From the lattice structure it is obvious that this connecting vector is one of the two primitive unit-cell vectors, and depending on the direction of the bond one needs to take $+1$ or $-1$ for $\nu_{ij}$. We will show during the discussion of the results that this choice is reasonable. 

\subsection{Two-band model and local SOC}
\label{subsec:two_band_model_and_local_soc}

Since the SOC term Eq.~(\ref{eq:bis_H_SOC}) in the two-band model is a local term, it suffices to look at the local self-energy. Although being local in terms of lattice coordinates, the local Hamiltonian as well as the local self-energy acquire spin-off-diagonal terms. The eigenbasis of the local non-interacting Hamiltonian Eq.~(\ref{eq:bis_H_0}) is given by the two basis functions
\begin{equation}
\label{eq:bis_eigenbasis_H_SO}
\begin{aligned}
&\left| p_{\pm}^{\uparrow} \right\rangle  = \frac{1}{\sqrt{2}}\left( \left| p_x^{\uparrow}\right\rangle  \pm i \left| p_y^{\uparrow}\right\rangle \right) \\
&\left| p_{\pm}^{\downarrow} \right\rangle  = \frac{1}{\sqrt{2}}\left( \left| p_x^{\downarrow}\right\rangle  \pm i \left| p_y^{\downarrow}\right\rangle \right), 
\end{aligned}
\end{equation}
where the eigenvalues are $\pm \lambda_{SO}^{2B}/2$, respectively. In this basis, the self-energy is diagonal. We denote the matrix elements by $\Sigma_{+/-}^{\sigma}$ with $\sigma = \uparrow,\downarrow$, which fulfil the relations $\Sigma_{+}^{\uparrow} = \Sigma_{-}^{\downarrow}$ and $\Sigma_{-}^{\uparrow} = \Sigma_{+}^{\downarrow}$. 
Similar to what has been done in three-band systems~\cite{triebl_spin-orbit_2018}, we can write the self-energy in this basis as
\begin{equation}
\Sigma = \frac{1}{2}\left( \Sigma_{+}^{\uparrow} + \Sigma_-^{\uparrow}\right) 1\!\!1 +  \left( \Sigma_{+}^{\uparrow} - \Sigma_-^{\uparrow}\right)\, \mathbf{l} \cdot \mathbf{s},
\end{equation}
where $1\!\!1$ is the identity and $\mathbf{l} \cdot \mathbf{s}$ is the SOC operator. 
From this relation an expression for an effective spin-orbit coupling can be defined as~\cite{triebl_spin-orbit_2018}
\begin{equation}
\label{eq:bis_lambda_eff}
\lambda_\text{eff}^{2B} = \lambda_{SO}^{2B} + \left( \text{Re}\,\Sigma_{+}^{\uparrow}(i\omega_0) - \text{Re}\,\Sigma_-^{\uparrow}(i\omega_0) \right).
\end{equation}

\section{TRILEX ($\Lambda^2$)}
\label{sec:trilex}

\begin{figure}[t]
\begin{center}
\includegraphics[width=\columnwidth]{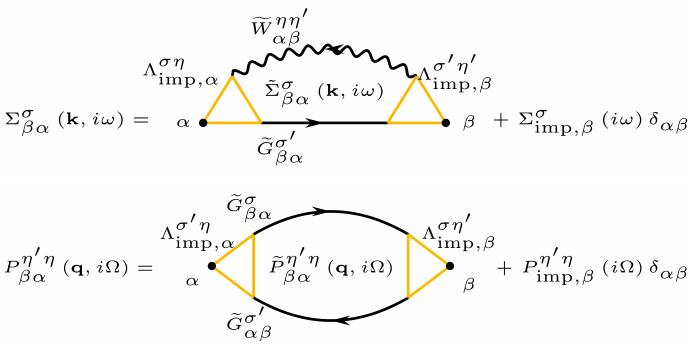}
\caption{Lattice self-energy $\Sigma(\textbf{k},i\omega)$ (upper panel) and polarization $P(\textbf{q},i\Omega)$ (lower panel) within TRILEX: the electron-boson coupling vertex $\Lambda$ is approximated by a local quantity in the Hedin equations, and $\alpha$ and $\beta$ denote sublattice indices.}
\label{fig:TRILEX_self_energies}
\end{center}
\end{figure}

The triply irreducible local expansion \cite{ayral_trilex_2015,ayral_trilex_2016} (TRILEX) is a diagrammatic extension of DMFT \citep{diagramatic_extensions_dmft} that was developed by Ayral and Parcollet and consists of decoupling the interaction term and approximating the electron-boson vertex function by a local one.
More precisely, in the Heisenberg decoupling, the interaction term is rewritten as
\begin{equation}
\label{eq:hubbard_interaction_decoupling}
U n_{i\uparrow} n_{i\downarrow} = \frac{1}{2}\sum_{I=0,x,y,z} U^{I} n^I_i n^I_i,
\end{equation}
with 
\begin{equation}
    \label{eq:decomposed_densities}
    n^I_i = \sum_{\sigma\sigma'}c^{\dagger}_{i\sigma}\sigma^{I}_{\sigma\sigma'}c_{i\sigma'},
\end{equation}
where $\sigma^0=\mathbf{1}$ and $\sigma^{x/y/z}$ are the Pauli matrices. $U^{I}$ denotes the bare interaction in the channel $I$. This decoupling, however, is not unique ("Fierz ambiguity") \cite{ayral_trilex_2015, ayral_fierz_trilex_2017}, and the ratio between the bare interactions $U^I$ in the different channels are determined by a parameter $\alpha$, the so-called Fierz parameter:
\begin{equation}
\label{eq:heisenberg_decoupling}
\begin{aligned}
U^{\text{ch}} &= U(3\alpha-1),\\
U^{x} = U^{y} = U^{z} &= U(\alpha-\frac{2}{3}).
\end{aligned}
\end{equation} 
Here we have definde the bare interaction in the charge-channel as $U^{\text{ch}} \equiv U^0$.
As the SOC term in the KMH model breaks SU(2) symmetry, we can assume spin-rotational invariance only in the xy-plane, leaving us with three independent channels: 
\begin{equation}
\label{eq:kmh_channels}
	\text{ch} = 0, \,xy = x/y, \, z=z
\end{equation}

Through Hubbard-Stratonovich transformations in each of the channels the electron-electron interaction problem is then transformed into an electron-boson coupling problem. The bosonic and fermionic lattice Green's functions are given by the Dyson equations
\begin{align}
G^{\sigma}(\mathbf{k},i\omega) &= [[G_0^{\sigma}(\mathbf{k},i\omega)]^{-1} -\Sigma^{\sigma}(\mathbf{k},i\omega)]^{-1}, \label{eq:dyson_fermions} \\
W^{\eta\eta'}(\mathbf{q},i\Omega) &= U^{\eta}[\delta_{\tilde{\eta}\tilde{\eta}'} - P^{\tilde{\eta}\tilde{\eta}'}(\mathbf{q},i\Omega)U^{\tilde{\eta}'}]_{\eta\eta'}^{-1}, \label{eq:dyson_bosons}
\end{align}
with $\eta \in \{ch, xy, z\}$, the momentum variables $\textbf{k}$ and $\textbf{q}$, the fermionic and bosonic Matsubara frequencies $\omega_n$ and $\Omega_n$, the non-interacting Green's function $G^{\sigma}_0(\mathbf{k},i\omega_n)$ and the polarization $P^{\eta\eta'}\left(\mathbf{q},i\Omega\right)$. Note that despite the spin-diagonal form of Green's function assumed here, the bosonic Green's functions will in general acquire channel off-diagonal components in the case of a broken SU(2) symmetry. 
The self-energy and polarization for a SU(2)-broken system are exactly given by the Hedin equations 
\begin{align}
P^{\eta'\eta}_{ji} &= \,\lambda^{\sigma'\eta}_{edi}G^{\sigma}_{ae}G^{\sigma'}_{db}\Lambda^{\sigma\eta'}_{baj}  \,\,\,\,  \sigma' = \begin{cases}
\bar{\sigma}, \,\,\text{if}\,\, \eta,\eta' \in [x,y]\\
\sigma, \,\,\text{if}\,\, \eta,\eta' \in [ch,z],
\end{cases} \label{eq:hedin_P_SU2_broken_main_text}\\
\Sigma^{\sigma}_{ji} &= -\lambda^{\sigma\eta}_{aib}G^{\sigma'}_{da}W^{\eta\eta'}_{bc}\Lambda^{\sigma'\eta'}_{jdc}
\,\,\,\,    \sigma' = \begin{cases}
\bar{\sigma}, \,\,\text{if}\,\, \eta,\eta' \in [x,y]\\
\sigma, \,\,\text{if}\,\, \eta,\eta' \in [ch,z],
\end{cases} \label{eq:hedin_Sigma_SU2_broken_main_text}
\end{align} 
where we have used roman literals for space-time indices and the Einstein convention for the summation over internal indices. $\lambda$ denotes the bare coupling vertices, which are defined via 
\begin{equation}
\label{eq:definition_bare_vertex}
     n^{\eta}_c\equiv c^\dagger_{a\sigma'}\lambda_{abc}^{\sigma\eta}c^{\phantom{\dagger}}_{b\sigma}.
\end{equation}
The TRILEX approximation consists in replacing the full electron-boson-coupling vertices $\Lambda^{\sigma\eta}_{abc}$ (c.f. Eq. \eqref{eq:Vertex_calculation_general}) by the ones of an effective impurity model $\Lambda^{\sigma\eta}_{\text{imp}}$. The original TRILEX approach however, due to its asymmetric diagrammatic structure, does not enforce hermiticity for channel-off-diagonal components of the polarization,
i.e. $P^{\eta\eta'}(\mathbf{q},i\Omega) = \left[P^{\eta'\eta}(\mathbf{q},-i\Omega)\right]^*$ is not automatically fulfilled. Therefore, we use a modified approach where the non-local self-energy and polarization is calculated from a Hedin diagram, where the renormalized electron-boson vertex of the impurity is inserted on both sides~\citep{double_Lambda1,double_Lambda2,double_Lambda3,double_Lambda4,dual_trilex1,dual_trilex2}. Following Ref.~\citep{double_Lambda4}, we will refer to this variant  as TRILEX $\Lambda^2$.
Simply speaking, we replace the bare vertex $\lambda$ entering Eqs.~(\ref{eq:hedin_P_SU2_broken_main_text}) and (\ref{eq:hedin_Sigma_SU2_broken_main_text}) by the renormalized electron-boson vertex $\Lambda$ while correcting for any double-counting of diagrams this entails. A more detailed derivation is given in Appendix~\ref{sec:derivation_of_the_su(2)_broken_trilex_equations}. 
In all equations in this paper we assume that all summations are properly normalized.  Summations over momenta $\sum_{\mathbf{k}/\mathbf{q}}$ include implicitly a prefactor $1/N_{\mathbf{k}/\mathbf{q}}$ with $N_{\mathbf{k}/\mathbf{q}}$ the number of $\mathbf{k}$-points in the Brillouin zone, and summations of frequencies $\sum_{i\omega/i\Omega}$ contain implicitly a prefactor $1/\beta$ with $\beta$ the inverse temperature.
The SU(2)-broken TRILEX $\Lambda^{2}$ approximation for a single-orbital model reads (cf. Fig. \ref{fig:TRILEX_self_energies} for a diagrammatic representation)
\begin{widetext}
\begin{align}
&\Sigma^{\sigma}_{\beta\alpha}(\mathbf{k},i\omega) = \underbrace{-\sum_{\mathbf{q}, i\Omega,\eta} m^{\eta}\Lambda^{\sigma\eta}_{\text{imp},\alpha}(i\omega+i\Omega,-i\Omega)\widetilde{G}^{\sigma'}_{\beta\alpha}(\mathbf{q + k},i\omega + i\Omega)\widetilde{W}^{\eta\eta'}_{\alpha\beta}(\mathbf{q},i\Omega)\Lambda^{\sigma'\eta'}_{\text{imp},\beta}(i\omega,i\Omega)}_{\widetilde{\Sigma}^{\sigma}_{\beta\alpha}(\mathbf{k},i\omega)} + \Sigma^{\sigma}_{\text{imp},\beta}(i\omega)\delta_{\alpha\beta}\label{eq:hedin_self_energy_TRILEX_approx}\\
&P^{\eta'\eta}_{\beta\alpha}(\mathbf{q},i\Omega) = \underbrace{\sum_{\mathbf{k}, i\omega,\sigma} \Lambda^{\sigma'\eta}_{\text{imp},\alpha}(i\omega+i\Omega,-i\Omega)\widetilde{G}^{\sigma}_{\beta\alpha}(\mathbf{q + k},i\omega + i\Omega)\widetilde{G}^{\sigma'}_{\alpha\beta}(\mathbf{k},i\omega)\Lambda^{\sigma\eta'}_{\text{imp},\beta}(i\omega,i\Omega)}_{\widetilde{P}^{\eta'\eta}_{\beta\alpha}(\mathbf{q},i\Omega)} + P^{\eta'\eta}_{\text{imp},\beta}(i\Omega)\delta_{\alpha\beta}\label{eq:hedin_polarization_TRILEX_approx}
\end{align}
with 
\begin{minipage}{.30\linewidth}
\begin{equation*}
m^{\eta} = 
\begin{cases}
2, \,\,\text{if}\,\, \eta = xy\\
1, \,\,\text{else}
\end{cases}  
\end{equation*}
\end{minipage}
\begin{minipage}{.30\linewidth}
\begin{equation*}
\sigma' = 
\begin{cases}
\bar{\sigma}, \,\,\text{if}\,\, \eta = xy\\
\sigma, \,\,\text{else}.
\end{cases} 
\end{equation*}
\end{minipage}
\end{widetext}

Here, $\Sigma_{\text{imp}}$ and $P_{\text{imp}}$ are the self-energy and polarization of the impurity, while $\alpha$ and $\beta$ denote sublattice indices (i.e., atoms in the unit cell). 
Quantities with a tilde in Eq.~\eqref{eq:hedin_self_energy_TRILEX_approx} and \eqref{eq:hedin_polarization_TRILEX_approx} are purely non-local, i.e.
\begin{equation}
    \label{eq:non-local_quantities_single_site}
    \widetilde{X}_{\alpha\beta}\left(\mathbf{k},i\omega\right) = X_{\alpha\beta}\left(\mathbf{k},i\omega\right) - \underbrace{\sum_{\mathbf{k}}X_{\alpha\alpha}\left(\mathbf{k},i\omega\right)}_{X_{\text{loc},\alpha}}\delta_{\alpha\beta}.
\end{equation}
In order to be consistent with our choice of the impurity, we have defined locality here on the level of the single atom, instead of the unit cell.
Note that here, different from the original TRILEX approximation~\citep{ayral_trilex_2015,ayral_trilex_2016}, the separation of \textit{local} (impurity) and  \textit{non-local} part of the self-energy and polarization is not just a numerical trick to improve on the convergence of frequency summations, but is necessary in order to avoid double counting of local diagrams. This also makes $\Sigma_{\text{loc}}$/$P_{\text{loc}}$  trivially equal to $\Sigma_{\text{imp}}$/$P_{\text{imp}}$.

In contrast to (E)DMFT, both the fermionic self-energy and the bosonic polarization, acquire a momentum dependence despite the local approximation of the electron-boson coupling vertex of TRILEX.
The expressions for the fermionic and bosonic Weiss fields $\mathcal{G^{\sigma}_{\alpha}}$ and $\mathcal{U^{\eta}_{\alpha}}$, as well as the impurity action $S_{\text{imp},\alpha}$, are identical to (E)DMFT: 
\begin{align}
\mathcal{G}^{\sigma}_{\alpha}(i\omega)& = \left[\left[G^{\sigma}_{\text{loc},\alpha}(i\omega)\right]^{-1} + \Sigma^{\sigma}_{\text{loc},\alpha}(i\omega)\right]^{-1}\label{eq:weiss_field_fermionic}\\
\mathcal{U}^{\eta\eta'}_{\alpha}(i\Omega)& = W^{\eta\eta'}_{\text{loc},\alpha}(i\Omega)\left[\delta_{\kappa\kappa'} - P^{\kappa\tilde{\kappa}}_{\text{loc},\alpha}(i\Omega)W^{\tilde{\kappa}\kappa'}_{\text{loc},\alpha}(i\Omega)\right]^{-1}_{\tilde{\eta}\eta'}\label{eq:weiss_field_bosonic}
\end{align}
\begin{equation}
\label{eq:impurity_action}
\begin{aligned}
S_{\text{imp},\alpha} = &-\int\int_0^{\beta}d\tau d\tau' \sum_{\sigma}c^{*}_{\sigma}(\tau)[\mathcal{G}_{\alpha}^{\sigma}(\tau-\tau')]^{-1}c_{\sigma}(\tau') 
\\&+ \int_0^{\beta}d\tau n_{\uparrow}(\tau)U n_{\downarrow}(\tau)
\\&+ 1/2 \int\int_0^{\beta}d\tau d\tau' \sum_{\sigma\sigma'}n_{\sigma}(\tau)\mathcal{D}_{\alpha}^{\sigma\sigma'}(\tau-\tau')n_{\sigma'}(\tau')
\\&+ 1/2 \int\int_0^{\beta}d\tau d\tau' s_{+}(\tau)\mathcal{J}_{\perp,\alpha}(\tau-\tau')s_{-}(\tau'),
\end{aligned}
\end{equation}
where we have defined
\begin{align}
    \mathcal{D}_{\alpha}^{\sigma\sigma'}(\tau) &\equiv \left[\mathcal{U}_{\alpha}^{\text{ch}}(\tau) - U^{\text{ch}}\right]+(-)^{\sigma\sigma'} \left[\mathcal{U}_{\alpha}^{z}(\tau) - U^{z}\right]\nonumber\\
    &+ (-)^{\sigma'} \mathcal{U}_{\alpha}^{\text{ch}, z} + (-)^{\sigma} \mathcal{U}_{\alpha}^{z, \text{ch}}\label{eq:dyniamical_interaction_D_sigma_sigma'}\\ 
    \mathcal{J}_{\perp,\alpha}(\tau) &\equiv 4\left[\mathcal{U}_{\alpha}^{xy}(\tau)-U^{xy}\right] \nonumber\\ 
    &+ 2i\left[ \mathcal{U}_{\alpha}^{x,y}(\tau) - \mathcal{U}_{\alpha}^{y,x}(\tau)\right] \label{eq:dyniamical_interaction_J_perpendicular'}\\
    s_{\pm} &= (n^{x} \pm in^{y})/2,
\end{align}
using the convention
\begin{align*}
    &(-)^{\sigma\sigma'} = \begin{cases} +1 \,\,\text{if}\,\, \sigma =\sigma' \\-1 \,\,\text{if}\,\, \sigma \neq\sigma' \end{cases}\\
    &(-)^{\uparrow} = +1,\,\,\, (-)^{\downarrow} = -1.
\end{align*}
    Note the difference between $\mathcal{U}^{xy}$, which denotes the two equivalent channel-diagonal elements $\mathcal{U}^{x}$ and $\mathcal{U}^{y}$, and $\mathcal{U}^{x,y}$ and $\mathcal{U}^{y,x}$, which denote channel-off-diagonal contributions. In Eq.~\eqref{eq:impurity_action} we have separated the bare Hubbard interaction $U$ from the dynamical $\mathcal{D}_{\alpha}^{\sigma\sigma'}(\tau)$.
    
The TRILEX self-consistency loop now consists of the following steps:
\begin{enumerate}[label=\textbf{TR\arabic*}]
    \item Choose a suitable \textbf{initial} \textbf{self-energy} $\Sigma^{\sigma}(\mathbf{k},i\omega)$ and \textbf{polarization} $P^{\eta}(\mathbf{q},i\Omega)$.
    \item Use the \textbf{Dyson equations} to calculate the re-normalized fermionic and bosonic lattice Green's functions (Eqs.~\eqref{eq:dyson_fermions} and \eqref{eq:dyson_bosons}).\label{item:TRILEX_iteration_dyson_equation}
    \item Calculate the \textbf{Weiss fields} from the local quantities (Eq.~\eqref{eq:weiss_field_fermionic} and \eqref{eq:weiss_field_bosonic}). \label{item:TRILEX_iteration_weiss_fields}
    \item Solve the \textbf{impurity models} (for every atom) using the impurity action in Eq.~\eqref{eq:impurity_action}. Calculate $\Sigma_{\text{imp}}$, $P_{\text{imp}}$ and $\Lambda_{\text{imp}}$.
    \item \textbf{Calculate} the \textbf{lattice self-energy} and \textbf{polarization} via the Hedin Eqs.~\eqref{eq:hedin_self_energy_TRILEX_approx} and \eqref{eq:hedin_polarization_TRILEX_approx}.\label{item:TRILEX_iteration_hedin_equations}
    \item \textbf{Inner TRILEX self-consistency loop} by repeatedly applying \ref{item:TRILEX_iteration_dyson_equation} and \ref{item:TRILEX_iteration_hedin_equations} (until convergence). \label{item:TRILEX_inner_loop}
    \item Go \textbf{back to} \ref{item:TRILEX_iteration_dyson_equation}.
\end{enumerate}
Note that \ref{item:TRILEX_inner_loop} is optional. In the highly correlated regime this step however helped stabilizing a paramagnetic solution.\\
For a more detailed discussion the reader is referred to \cite{ayral_trilex_2015,ayral_trilex_2016}.

In order to solve the paramagnetic KMH model we will also omit a broken sublattice symmetry by solving just one impurity problem, setting the individual impurity quantities equal, i.e.
\begin{align}
    \Lambda^{\sigma\eta}_{\text{imp},\alpha} &= \Lambda^{\sigma\eta}_{\text{imp}}\label{eq:single_impurity_vertex}\\
    \Sigma^{\sigma}_{\text{imp},\alpha} &= \Sigma^{\sigma}_{\text{imp}}\label{eq:single_impurity_self_energy}\\
    P^{\eta'\eta}_{\text{imp},\alpha} &= P^{\eta'\eta}_{\text{imp}}\label{eq:single_imurity_polarization}.
\end{align}
Enforcing paramagnetism then means
\begin{align}
\sigma^{\eta}_{\sigma\uparrow}\Lambda^{\uparrow\eta }_{\text{imp}} &\overset{!}{=} \sigma^{\eta}_{\sigma\downarrow}\Lambda^{\downarrow\eta }_{\text{imp}}\label{eq:kmh_lambda_imp_paramagnetic}\\
\Sigma^{\uparrow}_{\text{imp}} &\overset{!}{=} \Sigma^{\downarrow}_{\text{imp}} \equiv \Sigma_{\text{imp}}\label{eq:kmh_sigma_imp_paramagnetic}\\
    P^{\eta'\eta}_{\text{imp}} &= P^{\eta}_{\text{imp}}\delta_{\eta'\eta}\label{eq:kmh_p_imp_paramagnetic}.
\end{align}
For more details on why Eqs.~\eqref{eq:kmh_lambda_imp_paramagnetic} and \eqref{eq:kmh_p_imp_paramagnetic} hold in the paramagnetic case see Appendix~\ref{sec:implementation_trilex_kmh} and \ref{sec:channel_contributions_to_bosonic_gfs}.
The self-energy and polarization in our approximation therefore read
\begin{equation}
\label{eq:kmh_Sigma_approximation}
\Sigma^{\sigma} = \begin{pmatrix}
\tilde{\Sigma}^{\sigma}_{AA} + \Sigma_{\text{imp}}  & \Sigma_{AB}\\
\Sigma_{BA} & \tilde{\Sigma}^{\sigma}_{BB} + \Sigma_{\text{imp}}
\end{pmatrix}
\end{equation}

\begin{equation}
\label{eq:kmh_P_approximation}
P^{\eta\eta'} = \begin{pmatrix}
\tilde{P}_{AA}^{\eta\eta'} + P^{\eta}_{\text{imp}}\delta_{\eta\eta'}  & P^{\eta\eta'}_{AB}\\
P^{\eta\eta'}_{BA} & \tilde{P}_{BB}^{\eta\eta'} + \tilde{P}^{\eta}_{\text{imp}}\delta_{\eta\eta'}
\end{pmatrix}.
\end{equation}
In Appendix~\ref{sec:channel_contributions_to_bosonic_gfs} we also show that $P^{\text{ch},z}$ ($P^{z,\text{ch}}$) is the only non-zero off-diagonal contribution to the KMH model, when calculations are performed at half-filling, where the system is particle-hole symmetric \citep{zheng_particle-hole_2011}.

Calculations have been performed using a $30\times30$ $\mathbf{k}$-point grid.
The TRILEX ($\Lambda^2$) code has been implemented using the TRIQS library \citep{triqs}, and uses an implementation of Rubtsov's interaction expansion continuous-time quantum Monte-Carlo algorithm \citep{CTINT} as an impurity solver.

In~\citep{dual_trilex2} also paramagnetic calculations are performed on a model with non-local SOC using a dual-boson based approach. Quite generally, the TRILEX $\Lambda^2$ method presented here is related to other beyond-DMFT schemes such as the dual-boson~\cite{rubtsov_dual_2012_dual_boson1,loon_beyond_2014_dual_boson2} or the D-TRILEX approach~\cite{double_Lambda2,dual_trilex1,stepanov_orbital_2021_arxiv}. In all these approaches a double insertion of the vertex is performed. In the dual-boson schemes, since local and non-local contributions are separated, there is no problem with double counting of diagrams. The double insertion of the vertex just leads to more diagrams that are taken into account.
Let us stress here that dual-boson methods have been designed to target non-local interactions, while in our work we are restricting ourselves to local interactions only.

\subsection{Determination of the $\alpha$-parameter}
\label{subsec:fixing_the_alpha-parameter}
The choice of the Fierz parameter $\alpha$ in Eq.~\eqref{eq:heisenberg_decoupling} is influencing the results both at the one- and two-particle level \citep{ayral_trilex_2015,schaefer_toschi_2021}. This so called Fierz-ambiguity can be systematically reduced when the electron-boson vertex is extracted from clusters of increasing size instead of single-site impurities \citep{ayral_fierz_trilex_2017}, however at significantly increased computational costs.

For the two dimensional square lattice close to the antiferromagnetic (AF) phase transition (at $T=0$) spin-fluctuations are dominant \citep{Gunnarsson_fluctuation_diagnostics}. As we are enforcing a paramagnetic solution also in an AF regime, all calculations in the main text will refer to $\alpha=1/3$~\citep{schaefer_toschi_2021}, unless stated otherwise. In order to demonstrate that our conclusions do not depend on the choice of $\alpha$, in Appendix~\ref{sec:some_more_details_on_the_self_energy} we show results for a wide range of Fierz parameters.

\section{Results}
\begin{figure}
	\begin{center}
		\includegraphics[width=\columnwidth]{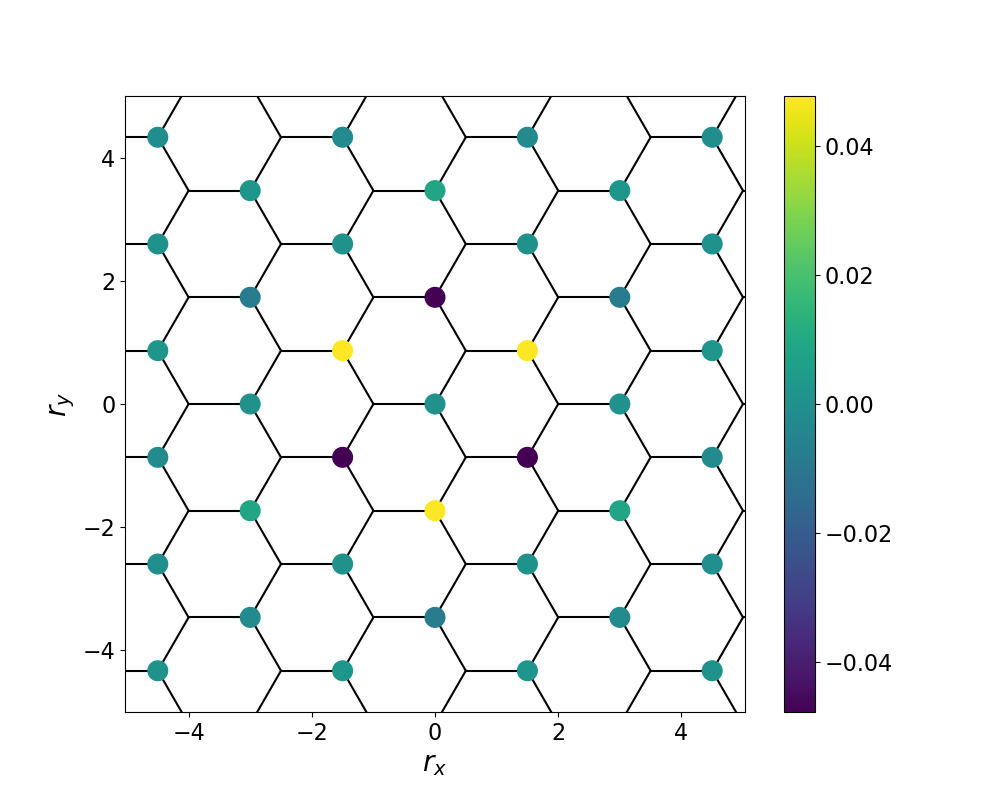}
		\caption{Illustration of $\text{Im}\, \Sigma^{SO}(\mathbf{r}-\mathbf{r}',i\omega_0)$. Predominantly next-nearest neighbors are contributing and right and left turns are associated with opposite signs (compare Fig.~\ref{fig:kmh_complex_hopping}). $U=7.0t$ and $\alpha=0.33$ }
		\label{fig:kmh_sigma_soc_real_space}
	\end{center}
\end{figure}

\begin{figure}
	\begin{center}
		\includegraphics[width=\columnwidth]{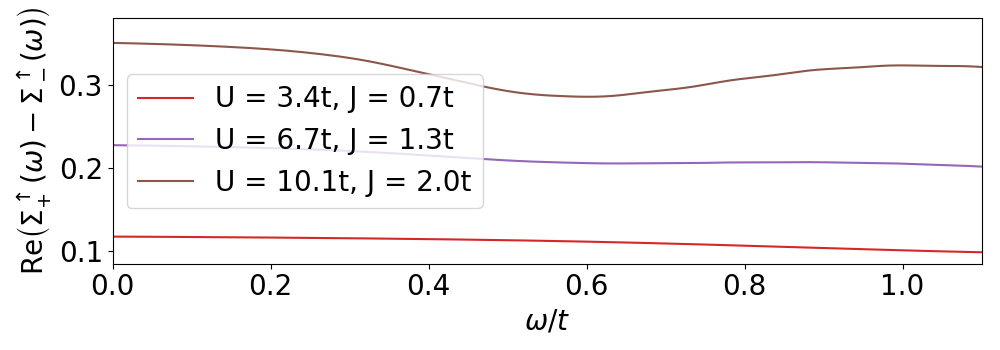}
		\includegraphics[width=\columnwidth]{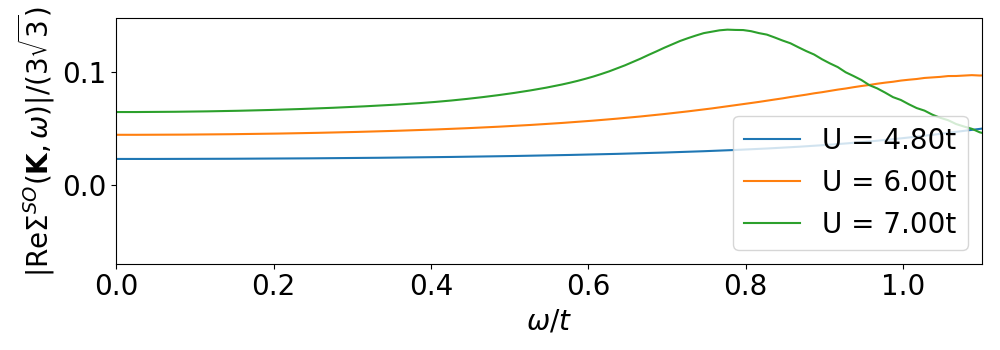}
		\caption{Frequency dependence of SOC-enhancement of the two-band model (upper panel) and the one-band model (lower panel) for different values of $U$ (and $J$).}
		\label{fig:kmh_bis_soc_eff_real_freq}
	\end{center}
\end{figure}

To show that local interactions increase the effective spin-orbit coupling in the KMH model, we first take a look at the structure of $\Sigma^{SO}$ and see if it matches the structure of the spin-orbit term of the Hamiltonian. Therefore, we perform a Fourier transformation to real space and compare $\textrm{Im}\,\Sigma^{SO}(\mathbf{r},i\omega_0)$ to the SOC term in Eq.~\eqref{eq:kmh_Hamiltonian}. From Fig.~\ref{fig:kmh_sigma_soc_real_space} we see that just like in Eq.~\eqref{eq:kmh_Hamiltonian} predominantly nearest-neighbor terms contribute to $\Sigma^{SO}$, and that right turns pick up a different sign than left turns. The real part turns out to be negligible.  This relationship also holds for higher 
Matsubara frequencies, i.e.,
\begin{subequations}
\begin{align}
&\text{Im}\,\Sigma^{SO}(\mathbf{r}-\mathbf{r}' = \mathbf{a}_{ij}, i\omega) \propto \nu_{ij}\lambda_{SO}^{1B} \,\,\,\,\, \forall\, i\omega,\label{subeq:kmh_prove_sigma_so_structure_a}\\
&\left|\frac{\text{Im}\,\Sigma^{SO}(\mathbf{r}-\mathbf{r}' \neq \mathbf{a}_{ij}, i\omega)}{\text{Im}\,\Sigma^{SO}(\mathbf{r}-\mathbf{r}' = \mathbf{a}_{ij}, i\omega)}\right| <  0.2\,\,\,\,\, \forall\, i\omega, \label{subeq:kmh_prove_sigma_so_structure_b}\\
&\left|\frac{\text{Re}\,\Sigma^{SO}(\mathbf{r}-\mathbf{r}', i\omega)}{\text{Im}\,\Sigma^{SO}(\mathbf{r}-\mathbf{r}' = \mathbf{a}_{ij}, i\omega)}\right| < 2\cdot10^{-3}\,\,\,\,\, \forall\, i\omega,\label{subeq:kmh_prove_sigma_so_structure_c}
\end{align}
\end{subequations}
with $\mathbf{a}_{ij}$ being a primitive unit vector. This shows that $\Sigma^{SO}$ has the same structure as the SOC term in Eq.~\eqref{eq:kmh_Hamiltonian} (see also Fig.~\ref{fig:kmh_complex_hopping}).
For the analytic continuation of the self-energy, we take its representation in $\mathbf{k}$-space. 
According to the correspondence of $\Sigma^{SO}$ to the complex next-nearest neighbor hopping term in real-space (Eq. \eqref{subeq:kmh_prove_sigma_so_structure_a} to \eqref{subeq:kmh_prove_sigma_so_structure_c}), in $\mathbf{k}$-space we find
\begin{subequations}
\begin{align}
    &\text{Re}\,\Sigma^{SO}\left(\mathbf{k},i\omega\right) \appropto \gamma_{\mathbf{k}}\,\,\,\,\, \forall\, i\omega, \label{subeq:kmh_sigma_so_structure_k_space_a}\\
    &\left|\text{Im}\,\Sigma^{SO}\left(\mathbf{k},i\omega\right)\right| \ll \left|\text{Re}\,\Sigma^{SO}\left(\mathbf{k},i\omega\right)\right|\,\,\,\,\, \forall\, i\omega. \label{subeq:kmh_sigma_so_structure_k_space_b}
\end{align}
\end{subequations}
From Eq.~\eqref{eq:kmh_gamma_k} we see that $\gamma_{\mathbf{k = K/K'}} = \mp3\sqrt{3}\lambda^{1B}_{SO}$ and, hence, together with Eq.~\eqref{subeq:kmh_prove_sigma_so_structure_a} we find
\begin{equation}
    \label{eq:SOC_enhancement_K_point}
    \nu_{ij}\text{Im}\,\Sigma^{SO}(\mathbf{r}-\mathbf{r}' = \mathbf{a}_{ij}, i\omega) \approx \frac{\text{Re}\,\Sigma^{SO}\left(\mathbf{k=K'},i\omega\right)}{3\sqrt{3}}.
\end{equation}
Note that this relation becomes exact if the terms in Eq.~\eqref{subeq:kmh_prove_sigma_so_structure_b} and \eqref{subeq:kmh_prove_sigma_so_structure_c} completely vanish. We can thus analytically continue the right-hand side of Eq.~\eqref{eq:SOC_enhancement_K_point} to get the frequency-dependent SOC enhancement~\footnote{For stability reasons we analytically continue $\Sigma^d(\mathbf{K'},i\omega)+\Sigma^{SO}(\mathbf{K'},i\omega)$ and $\Sigma^d(\mathbf{K'},i\omega)-\Sigma^{SO}(\mathbf{K'},i\omega)$ instead of $\Sigma^{SO}(\mathbf{K'},\omega)$ directly.}.
We perform this analytic continuation using the maximum entropy method implemented in the TRIQS library \citep{PhysRevB.96.155128,MaxEnt_kraberger}. The result is shown in the lower panel of Fig.~\ref{fig:kmh_bis_soc_eff_real_freq} for several different values of $U$, and we find that at least for small frequencies the SOC-enhancement is a constant, making our definition of $\lambda^{1B}_{\text{eff}}$ in Eq.~\eqref{eq:kmh_lambda_eff} reasonable. 

\begin{figure}
	\begin{center}
		\includegraphics[width=\columnwidth]{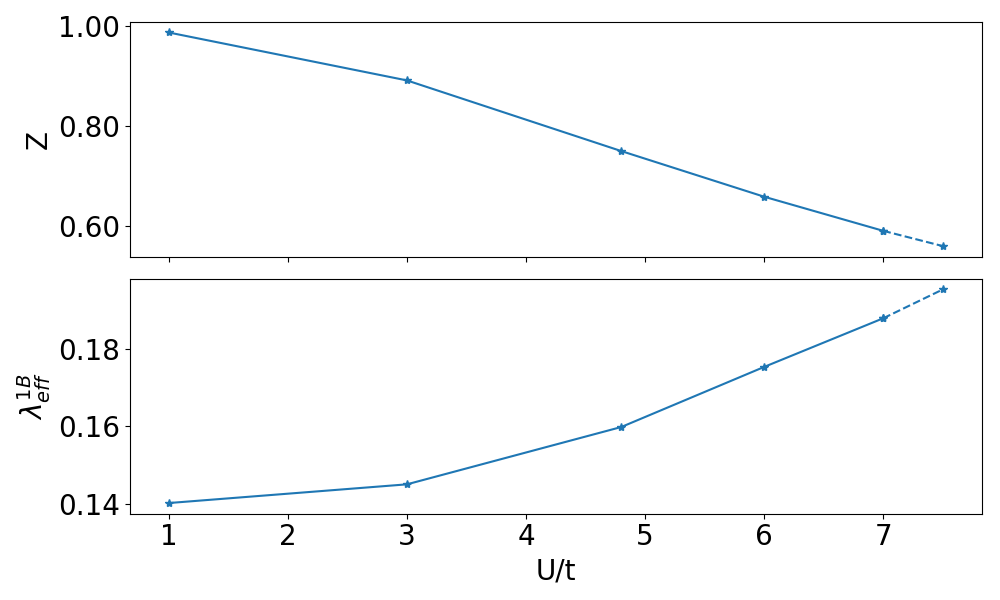}
		\caption{Evolution of the $\lambda_\text{eff}^{1B}/t$ and the renormalization factor $Z$ with increasing interaction $U/t$. The dashed line denotes a regime where it becomes increasingly hard to stabilize a paramagnetic solution.}
		\label{fig:kmh_z_and_lambda_eff}
	\end{center}
\end{figure}

In Fig.~\ref{fig:kmh_z_and_lambda_eff} we show how $\lambda_{\text{eff}}^{1B}$ develops with increasing on-site interaction $U/t$.  In order to quantify the correlation strength accompanied by the different values of $U/t$, the renormalization factor
\begin{equation}
\label{eq:quasi-particle_weight}
Z = \lim_{\omega \rightarrow 0} \left[1-\frac{\partial \text{Im}\,\Sigma_{\text{imp}}(i\omega)}{\partial \omega}\right]^{-1}
\end{equation}
is also shown in Fig.~\ref{fig:kmh_z_and_lambda_eff}. In order to approximate the derivative in Eq.~\eqref{eq:quasi-particle_weight} we use the slope between the two Matsubara frequencies  $\omega_0$ and $\omega_1$. This is not a real quasi-particle weight in the  sense of a Landau-liquid theorem (there are no states at the Fermi-level in the non-interacting model already, see Fig.~\ref{fig:Dispersion_comp}), however, this still quantifies the level of correlation in the system and is also bound between $1$ and $0$.
From Fig.~\ref{fig:kmh_z_and_lambda_eff} we see how, while correlations are getting stronger ($Z$ is dropping), $\lambda_\text{eff}^{1B}$ increases gently. 

\begin{figure}
	\begin{center}
	\includegraphics[width=\columnwidth]{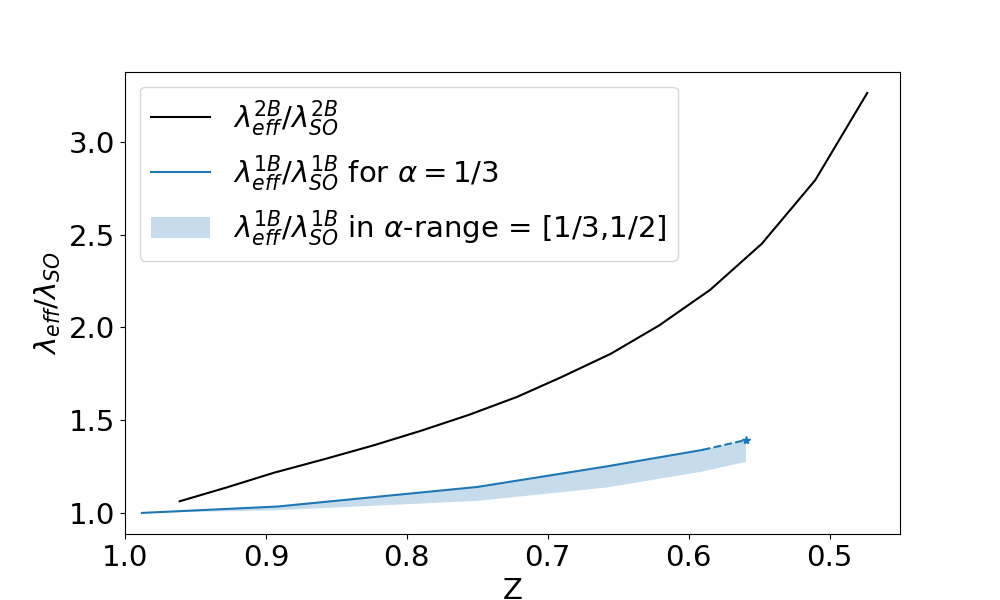}
		\caption{Compare the increase of $\lambda_{\text{eff}}$ with respect to a decreasing renormalization factor $Z$ for both models.}
		\label{fig:kmh_bis_lambda_eff_compared}
	\end{center}
\end{figure}

At this point we want to emphasize again that we are performing paramagnetic calculations by symmetrizing the impurity quantities. Without symmetrization the system would order above a critical $U^{c(AF)}$ between $4t-5t$, depending on the underlying method. In lattice quantum Monte Carlo a value around $U^{c}\gtrsim 5t$ \cite{hohenadler_quantum_KMH_2012}, and for cluster-DMFT $U^{c} \sim 5t$ \cite{wu_qsh_2012} was found. The authors find $U^{c} \gtrsim 4t$ for single-site DMFT. This means that half of the plot in Fig.~\ref{fig:kmh_z_and_lambda_eff} belongs to a regime that has a high tendency towards ordering. We had to use a quite low mixing factor (between 0.3 and 0.15) of the Weiss fields (\ref{item:TRILEX_iteration_weiss_fields}) but also of the lattice self-energy and polarization in the inner TRILEX self-consistency loop (\ref{item:TRILEX_inner_loop}) in order to stabilize paramagnetic solutions. Calculations with $U=7.5t$ have become already quite unstable (denoted by the dashed line in Fig.~\ref{fig:kmh_z_and_lambda_eff} and \ref{fig:kmh_bis_lambda_eff_compared}) and we have not been able to stabilize solutions for larger values of $U$. Note however that enforcing the paramagnetic solution is important in order to make the one-band model comparable to the two-band model.

\begin{figure}
	\begin{center}
	\includegraphics[width=\columnwidth]{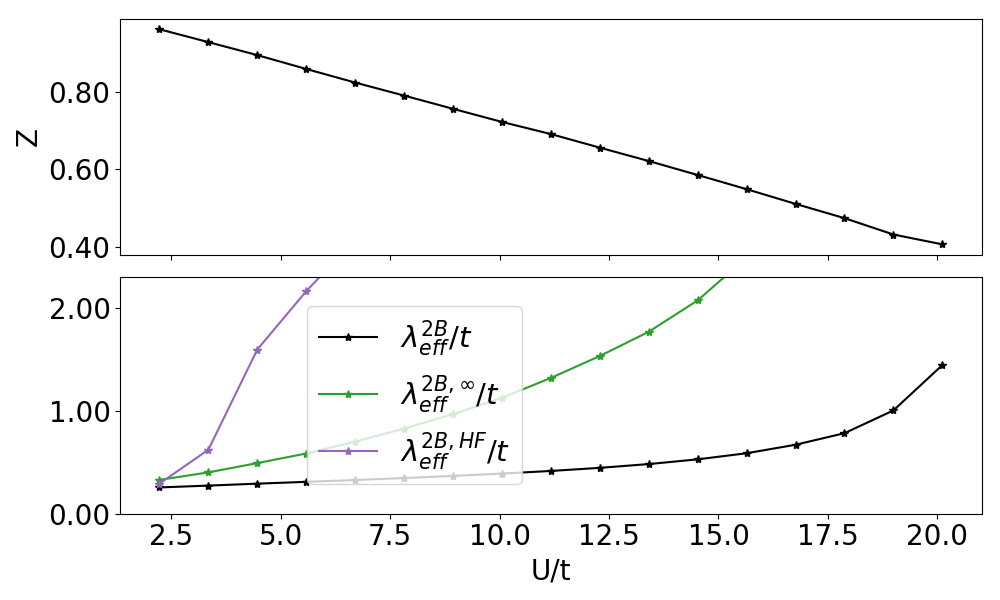}
		\caption{Evolution of the renormalization factor $Z$ and $\lambda_{\text{eff}}^{2B}$/$\lambda_{\text{eff}}^{2B,\infty}$/$\lambda_{\text{eff}}^{2B,HF}$ with increasing interaction $U$. $\lambda_{\text{eff}}^{2B,HF}$ is the self-consistent Hartree-Fock solution calculated from Eq. \eqref{eq:hartree_2B_model} and \eqref{eq:hartree_fock_interaction}. $\lambda_{\text{eff}}^{2B,\infty}$ denotes the high-frequency contribution to the effective SOC and is given by Eq. \eqref{eq:hartree_2B_model} when the densities are taken from DMFT (Eq. \eqref{eq:lambda_enhancement_high_frequency}).  }
		\label{fig:bis_z_and_lambda_eff}
	\end{center}
\end{figure}

For the two-band model of bismuthene we also look at the real-frequency dependence of the SOC-enhancement, again by applying the maximum entropy method. The top panel of Fig.~\ref{fig:kmh_bis_soc_eff_real_freq} shows that also here the enhancement is roughly constant in frequency. Unlike the one-band model this holds even at large frequencies. Hence the definition of $\lambda_\text{eff}^{2B}$ in Eq.~\eqref{eq:bis_lambda_eff} is also plausible. Fig.~\ref{fig:bis_z_and_lambda_eff} shows that for bismuthene $Z$ decreases quite linearly with $U/t$, while $\lambda_{\text{eff}}^{2B}$ starts diverging.

In order to compare the one-band and two-band model, Fig.~\ref{fig:kmh_bis_lambda_eff_compared} shows for both models the relative increment of SOC $\lambda_{\text{eff}}/\lambda_{SO}$ as a function of the renormalization factor $Z$. Here we have further added results for additional $\alpha$-parameters in the interval $[1/3,1/2]$, showing that results remain qualitatively consistent over the full range of $\alpha$-values (c.f. also Appendix~\ref{subsec:alpha-dependence_of_lambda_eff_1B}). We find that the enhancement of \textit{non-local} SOC in the one-band model is negligible compared to the enhancement of \textit{local} SOC in the two-band model.

The reason for this different response of the effective SOC to interactions can be traced back at least qualitatively to the specific forms of the SOC Hamiltonians as compared to the interaction Hamiltonians. For the one-band model with a non-local SOC Hamiltonian, the effective SOC is defined from a \textit{non-local} self-energy, while the microscopic Hubbard interaction is purely \textit{local}. For the two-band model, however, where the relevant self-energy component for the effective SOC is a local but orbital off-diagonal one, the local microscopic Kanamori interaction includes corresponding inter-orbital interaction terms. Thus, the interaction Hamiltonian is directly effecting the aforementioned self-energy terms. We believe this discrepancy between the locality of the SOC and the microscopic interaction to be the reason for the stronger response in the two-band case as opposed to the one-band case. In order to corroborate this assumption, we can treat the effective enhancement of SOC in Hartree-Fock (HF) theory. As we detail in Appendix \ref{sec:simpler_approximations}, the HF decoupling in the two-band case leads to a contribution that is exactly of the form of the SOC Hamiltonian. For the one-band case, such a contribution is absent by construction. 

We can also quantify the HF estimate for the SOC enhancement. First, we evaluate the SOC formula for the enhancement (Eq. \eqref{eq:hartree_2B_model}), using the converged DMFT results for the spin-orbital polarization, which is shown as green line in the lower panel of Fig.~\ref{fig:bis_z_and_lambda_eff}. Second, we also solved the HF equations self-consistently, the result of which is shown as the purple line in the above mentioned figure. In general, both HF estimates overestimate the DMFT results, where the self-consistent solution shows particularly strong enhancement. The reason is that in HF, already a rather small interaction $U$ and $J$ is sufficient to give an almost saturated polarization, and in turn strong enhancement. This is consistent with the considerations in Ref.~\cite{liu_coulomb-enhanced_soc_2008}, where from an almost fully spin-orbital polarized system a rather small interaction $U-J=0.5$\,eV was deduced.
As a result, these HF estimates show that it is important to take the dynamical correlation effects in the system properly into account.

\section{Conclusions}

In this work we investigated the effective SOC strength in a one-band and a two-band honeycomb lattice model. The main difference between the models is the structure of the SOC coupling term, which in the one-band case is non-local as opposed to the purely local term in the two-band case. In order to quantify this effect, we have defined
an effective SOC $\lambda_{\text{eff}}^{1B/2B}$ (Eqs.~\eqref{eq:kmh_lambda_eff} and \eqref{eq:bis_lambda_eff}). Both increase with increasing interaction strength. However, the effective enhancement in the one-band case is rather minute. 
One main reason for this difference is that in the two-band model both the interactions and the SOC are local terms. We find that a Hartree-Fock treatment of the interactions already leads to a sizable enhancement of the SOC. This effect is completely absent in the one-band model. However, we have furthermore shown that the Hartree-Fock treatment alone drastically overestimates this enhancement.

For the two-band case, within the treatment of DMFT, we see strong enhancements up to factors of 3 when correlations become sizable. Thus, for a similar correlation strentgh as measured by the renormalization factor $Z$, the enhancement in the two-band model is approximately one order of magnitude larger than in the one-band model. This suggests that effective SOC is a relevant concept in multi-orbital systems, but might be of rather limited importance in single-band systems with non-local SOC, at least for the correlation regime that we studied here. This possibly changes in the vicinity of phase transitions, or when also non-local interactions are included into the KMH, as they may considerably enhance the non-local effective SOC.

On the technical level, we have formulated the necessary equations for TRILEX in the case when SU(2) symmetry is broken. Since the original TRILEX formulations breaks hermiticity of certain quantities, we adopted the so-called TRILEX $\Lambda^2$ method. This broken symmetry is not only relevant for the study that we presented here, but of course also for models with long-range magnetic ordering. The formalism, without the enforced paramagnetic properties as developed here, is directly applicable to the magnetic case.

\acknowledgments
We thank T.~Ayral and O.~Parcollet for insightful discussions. This work has been supported by the Austrian Science Fund (FWF), grant number Y746. A part of the calculations have been performed on the Vienna Scientific Cluster. 

\appendix

\section{Derivation of the single-orbital SU(2)-broken TRILEX equations}
\label{sec:derivation_of_the_su(2)_broken_trilex_equations}

\begin{figure}
	\begin{center}
		\includegraphics[width=\columnwidth]{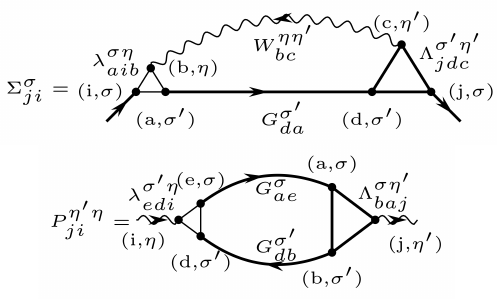}
		\caption{General SU(2)-broken Hedin-Equation of the self-energy (upper panel) and polarization (lower panel).}
		\label{fig:hedin_SU2_broken}
	\end{center}
\end{figure}

The most general formulation of the Hedin equations for the fermionic self-energy and the bosonic polarization reads \citep{ayral_trilex_2016}
\begin{equation}
\label{eq:hedin_general_eq}
\begin{aligned}
P^{\eta'\eta}_{ji} &= \lambda^{\tilde{\sigma}\sigma'\eta}_{edi}G^{\sigma\tilde{\sigma}}_{ae}G^{\sigma'\tilde{\sigma}'}_{db}\Lambda^{\tilde{\sigma}'\sigma\eta'}_{baj}\\
\Sigma^{\tilde{\sigma}\sigma}_{ji} &= -\lambda^{\tilde{\sigma}'\sigma\eta}_{aib}G^{\sigma'\tilde{\sigma}'}_{da}W^{\eta\eta'}_{bc}\Lambda^{\tilde{\sigma}\sigma'\eta'}_{jdc},
\end{aligned}
\end{equation}
where $a,b,c,d,e,f,i,j$ are space-time indices, $\sigma, \sigma', \tilde{\sigma}, \tilde{\sigma}'$ denote spin indices and $\eta,\eta'$ denote Pauli indices $\{0,x,y,z\}$  and Einstein's summation convention is adopted. $\lambda$ is the bare fermion-boson coupling vertex and $\Lambda$ it's fully renormalized version.\\
For the derivation of the equations for the specific purpose of this paper, we assume a spin-diagonal form of the fermionic Green's functions, i.e. $G^{\sigma\sigma'}_{ij} = G^{\sigma}_{ij}\delta_{\sigma\sigma'}$.\\
The polarization hence becomes
\begin{align}
	P^{\eta'\eta}_{ji} &= \lambda^{\sigma\sigma'\eta}_{edi}G^{\sigma}_{ae}G^{\sigma'}_{db}\Lambda^{\sigma'\sigma\eta'}_{baj}.
\end{align}
For the vertex $\Lambda^{\sigma'\sigma\eta'}_{baj}$ spin and channel indices are connected in the sense that $\Lambda^{\sigma'\sigma\eta'}_{baj} = \Lambda^{\sigma\eta'}_{baj}|\sigma^{\eta'}_{\sigma'\sigma}|$. Given that, we find 
\begin{equation}
\label{eq:hedin_P_SU2_broken}
\begin{aligned}
P^{\eta'\eta}_{ji} &= \lambda^{\sigma'\eta}_{edi}G^{\sigma}_{ae}G^{\sigma'}_{db}\Lambda^{\sigma\eta'}_{baj}  \,\,\,\,\,\,\,    \sigma' = \begin{cases}
\bar{\sigma}, \,\,\text{if}\,\, \eta,\eta' \in [x,y]\\
\sigma, \,\,\text{if}\,\, \eta,\eta' \in [ch,z],
\end{cases} 
\end{aligned}
\end{equation}
which is Eq.~\eqref{eq:hedin_P_SU2_broken_main_text} of the main text.
This gives, beside, the diagonal terms $P^{\eta\eta}_{ji}$, also 4 non-zero off-diagonal terms:
Using the same assumptions as above for the self-energy we find
\begin{equation}
\label{eq:hedin_sigma_simplified}
\Sigma^{\tilde{\sigma}\sigma}_{ji} = -\lambda^{\sigma'\sigma\eta}_{aib}G^{\sigma'}_{da}W^{\eta\eta'}_{bc}\Lambda^{\tilde{\sigma}\sigma'\eta'}_{jdc}
\end{equation}
and hence, by explicitly writing the vertex
\begin{align}
\Sigma^{\tilde{\sigma}\sigma}_{ji} = -\lambda^{\sigma\eta}_{aib}G^{\sigma'}_{da}W^{\eta\eta'}_{bc}\Lambda^{\sigma'\eta'}_{jdc}|\sigma^{\eta}_{\sigma'\sigma}||\sigma^{\eta'}_{\tilde{\sigma}\sigma'}|.
\end{align}
Given a density-density interaction as presented in Eq.~\eqref{eq:hubbard_interaction_decoupling}, the bare part of $W^{\eta\eta'}$ will be fully channel-diagonal, and channel-offdiagonal components get generated only through the polarization (see Eq.~\eqref{eq:dyson_bosons}). As $P$ is block-diagonal ($[ch,z]$ and $[x,y]$) the restrictions of Eq.~\eqref{eq:hedin_P_SU2_broken} pass on to $W^{\eta\eta'}$. Hence we only get non-zero contributions to the self-energy if $\sigma = \tilde{\sigma}$, leaving $\Sigma_{ij}$ spin-diagonal. The equation simplifies to 
\begin{align}
\label{eq:hedin_Sigma_SU2_broken}
\Sigma^{\sigma}_{ji} = -\lambda^{\sigma\eta}_{aib}G^{\sigma'}_{da}W^{\eta\eta'}_{bc}\Lambda^{\sigma'\eta'}_{jdc}
\,\,\,\,\,\,\,    \sigma' = \begin{cases}
\bar{\sigma}, \,\,\text{if}\,\, \eta,\eta' \in [x,y]\\
\sigma, \,\,\text{if}\,\, \eta,\eta' \in [ch,z],
\end{cases} 
\end{align}
which is Eq.~\eqref{eq:hedin_Sigma_SU2_broken_main_text} of the main text.

The graphical representations of Eqs.~\eqref{eq:hedin_P_SU2_broken} and \eqref{eq:hedin_Sigma_SU2_broken} can be found in Fig.~\ref{fig:hedin_SU2_broken}.

\subsection{Local vertex approximation}
\subsubsection{TRILEX}
Now we will use the fact that we are only considering Hubbard-type interactions. Therefore the bare vertex can be rewritten as $\lambda_{aib}^{\sigma\eta} = \sigma^{\eta}_{\sigma'\sigma}\delta_{ai}\delta_{ib}$ and Eq.~\eqref{eq:hedin_Sigma_SU2_broken} becomes
\begin{align}
\label{eq:hedin_Sigma_SU2_broken_hubbard_type}
\Sigma^{\sigma}_{ji} = -\sigma^{\eta}_{\sigma'\sigma}G^{\sigma'}_{di}W^{\eta\eta'}_{ic}\Lambda^{\sigma'\eta'}_{jdc}.
\end{align}

In the following $\alpha, \beta, \gamma, \delta$ denote inner unit cell indices (e.g. $A, B$) and $\mathbf{R},\mathbf{R}_1, \mathbf{R}_2$ are lattice vectors connecting unit cells. By omitting the spin/channel indices for clarity Eq.~\eqref{eq:hedin_Sigma_SU2_broken} becomes:

\begin{widetext}

\begin{equation}
\Sigma_{\beta\alpha}(\mathbf{R},\tau) \equiv -\int_{\tau_1 \tau_2} \sum_{\mathbf{R}_1,\mathbf{R}_1} \sum_{\gamma,\delta} G_{\gamma\alpha}(\mathbf{R}_1,\tau_1)W_{\alpha\delta}(-\mathbf{R}_2,-\tau_2)\Lambda_{\beta,\gamma,\delta}(\mathbf{R}-\mathbf{R}_1,\mathbf{R}_2-\mathbf{R}_1,\tau-\tau_1,\tau_2-\tau_1).
\end{equation}
Now, assuming the locality of the electron-boson coupling vertex
\begin{equation}
\label{eq:TRILEX_approximation}
\Lambda_{\beta,\delta,\gamma}(\mathbf{R}-\mathbf{R}_1,\mathbf{R}_2-\mathbf{R}_1,\tau-\tau_1,\tau_2-\tau_1) \approx \Lambda_{\beta}(\tau-\tau_1,\tau_2-\tau_1)\delta_{\beta\delta}\delta_{\beta\gamma}\delta_{\mathbf{R}\mathbf{R}_1}\delta_{\mathbf{R}_1\mathbf{R}_2},
\end{equation}
and approximating the local vertex by the impurity vertex (TRILEX approximation)
\begin{equation}
    \label{eq:vertex_approximation_impurity_vertex}
    \Lambda_{\beta} \approx \Lambda_{\text{imp},\beta},
\end{equation}
we obtain
\begin{equation}
\label{eq:hedin_TRILEX_space_time}
\Sigma_{\beta\alpha}(\mathbf{R},\tau) \equiv -\int_{\tau_1 \tau_2} G_{\beta\alpha}(\mathbf{R},\tau_1)W_{\alpha\beta}(-\mathbf{R},-\tau_2)\Lambda_{\text{imp},\beta}(\tau-\tau_1,\tau_2-\tau_1).
\end{equation}
Taking the following definition of the Fourier transform
\begin{equation}
X_{\alpha\beta}(\mathbf{k},i\omega) = \int_{\tau} e^{i\omega\tau} \sum_{\mathbf{R}_i} e^{i\mathbf{k}\mathbf{R}_i}X_{\alpha\beta}(\mathbf{R},\tau),
\end{equation}
we find by also adding back spin/channel indices
\begin{equation}
\label{eq:Sigma_Hedin_diagrams_reg}
\begin{aligned}	
\Sigma^{\sigma}_{\beta\alpha}(\mathbf{k},i\omega) = -\sum_{\mathbf{q}, i\Omega,\eta} &\sigma^{\eta}_{\sigma'\sigma}{G}^{\sigma'}_{\beta\alpha}(\mathbf{q + k},i\omega + i\Omega){W}^{\eta\eta'}_{\alpha\beta}(\mathbf{k},i\omega)\Lambda^{\sigma'\eta'}_{\text{imp},\beta}(i\omega,i\Omega)
\end{aligned}
\end{equation}
and equivalently
\begin{equation}
\label{eq:P_Hedin_diagrams_reg}
\begin{aligned}	
P^{\eta'\eta}_{\beta\alpha}(\mathbf{q},i\Omega) = \sum_{\mathbf{k}, i\omega,\sigma} &\sigma^{\eta}_{\sigma\sigma'}{G}^{\sigma}_{\beta\alpha}(\mathbf{q + k},i\omega + i\Omega){G}^{\sigma'}_{\alpha\beta}(\mathbf{k},i\omega)\Lambda^{\sigma\eta'}_{\text{imp},\beta}(i\omega,i\Omega).
\end{aligned}
\end{equation}
This formulation does not enforce the required Hermitian symmetry for channel-off-diagonal elements namely  $P^{\eta\eta'}_{\alpha\beta}(\mathbf{q},i\Omega) = \left[P^{\eta'\eta}_{\beta\alpha}((\mathbf{q},-i\Omega)\right]^*$. In order to stay consistent it is necessary to restrict the bosonic Green's functions to 
\begin{equation}
\label{eq:restriction_TRILEX}
    X^{\eta\eta'} \overset{!}{=} X^{\eta}\delta_{\eta\eta'}.
\end{equation}

\subsubsection{TRILEX $\Lambda^2$}

In order to enforce hermiticity of the polarization without requiring channel-diagonality we adopt the TRILEX $\Lambda^2$ approach,  where the non-local part of the self-energy $\widetilde{\Sigma}$ and polarization $\widetilde{P}$ is calculated from a diagram with two renormalized impurity-vertices appearing on both sides.\\
The non-local part of Eq.~\eqref{eq:hedin_TRILEX_space_time} hence becomes 
\begin{equation}
\widetilde{\Sigma}_{\beta\alpha}(\mathbf{R},\tau) = -\int_{\tau_1 \tau_2,\tau_3 \tau_4}\Lambda_{\text{imp},\alpha}(\tau_1,\tau_2) \widetilde{G}_{\beta\alpha}(\mathbf{R},\tau_3-\tau_1)\widetilde{W}_{\alpha,\beta}(-\mathbf{R},\tau_2-\tau_4)\Lambda_{\text{imp},\beta}(\tau-\tau_3,\tau_4-\tau_3).
\end{equation}
and after Fourier transform we obtain
\begin{equation}
\label{eq:hedin_self_energy_TRILEX_approx_appendix}
\widetilde{\Sigma}_{\beta\alpha}(\mathbf{k},i\omega) = -\sum_{\mathbf{q} \in RBZ, i\Omega}\Lambda_{\text{imp},\alpha}(i\omega+i\Omega,-i\Omega)\widetilde{G}_{\beta\alpha}(\mathbf{q + k},i\omega + i\Omega)\widetilde{W}_{\alpha\beta}(\mathbf{q},i\Omega)\Lambda_{\text{imp},\beta}(i\omega,i\Omega),
\end{equation}
and similarly
\begin{equation}
\label{eq:hedin_polarization_TRILEX_approx_appendix}
\widetilde{P}_{\beta\alpha}(\mathbf{q},i\Omega) = \sum_{\mathbf{q} \in RBZ, i\Omega}\Lambda_{\text{imp},\alpha}(i\omega+i\Omega,-i\Omega) \widetilde{G}_{\beta\alpha}(\mathbf{q + k},i\omega + i\Omega)\widetilde{G}_{\alpha\beta}(\mathbf{k},i\omega)\Lambda_{\text{imp},\beta}(i\omega,i\Omega).
\end{equation}
Recovering spin and channel indices and replacing the local quantities $\Sigma_{\text{loc}}$/$P_{\text{loc}}$ by the impurity ones $\Sigma_{\text{imp}}$/$P_{\text{imp}}$, leads Eq.~\eqref{eq:hedin_self_energy_TRILEX_approx} and \eqref{eq:hedin_polarization_TRILEX_approx} in the main text.

\subsection{Calculation of the Vertex}
The full electron-boson coupling vertex can be written as \citep{ayral_trilex_2016}
\begin{equation}
    \label{eq:Vertex_calculation_general}
    \Lambda^{\sigma'\sigma\eta}_{abc} = \left[G^{-1}\right]^{\sigma'\tilde{\sigma}'}_{ad}\left[G^{-1}\right]^{\tilde{\sigma}\sigma}_{eb}\left[\left(\delta_{gh}\delta_{\tilde{\eta}\tilde{\eta}'} - \chi^{\tilde{\eta}\tilde{\eta}'}_{gi}U^{\tilde{\eta}'}_{ih}\right)^{-1}\right]^{\eta\eta'}_{cf}\tilde{\chi}^{\tilde{\sigma}'\tilde{\sigma}\eta'}_{def},
\end{equation}
with the fermionic three-point correlation function 
\begin{equation}
    \label{eq:chi3_definition}
    \tilde{\chi}^{\tilde{\sigma}'\tilde{\sigma}\eta}_{def} = \left<c^{\tilde{\sigma}'}_{d}\bar{c}_{e}^{\tilde{\sigma}}\left(n_f^{\eta} - \left<n_f^{\eta}\right>\right)\right>,
\end{equation}
the susceptibility $\chi^{\eta\eta'}_{cf}$ (see Eq.~\eqref{eq:susceptibility}) and the decoupled interaction $U^{\eta}_{ih}$ (see Eq.~\eqref{eq:hubbard_interaction_decoupling}). 
Here again $a-i$ are space-time indices, $\bar{c}^{\tilde{\sigma}}_{d}$ ($c^{\tilde{\sigma}}_{d}$) are creation (annihilation) operators and $n^{\eta}_{g}$ is given by Eq.~\eqref{eq:decomposed_densities}.
Applying the same considerations as above and replacing the interaction by the according Weiss field (Eq.~\eqref{eq:weiss_field_bosonic}), the impurity vertex is given by 
\begin{equation}
    \label{eq:Impurity_vertex_general}
    \Lambda^{\sigma\eta}_{\text{imp},\alpha}\left(i\omega,i\Omega\right) = \left[G_{\text{imp},\alpha}^{\sigma'}\left(i\omega+i\Omega\right)\right]^{-1} \left[G_{\text{imp},\alpha}^{\sigma}\left(i\omega\right)\right]^{-1} \left[\left(\delta_{\kappa\kappa'} - \chi_{\text{imp},\alpha}^{\kappa\tilde{\kappa}}\left(i\Omega\right) \mathcal{U}^{\tilde{\kappa}\kappa'}_{\alpha}\left(i\Omega\right)\right)^{-1}\right]^{\eta\eta'}\tilde{\chi}_{\text{imp},\alpha}^{\sigma\eta'}\left(i\omega,i\Omega\right).
\end{equation}

\end{widetext}

\section{Implementation details for the paramagnetic KMH}
\label{sec:implementation_trilex_kmh}
Calculating the impurity vertex for a single-site paramagnetic calculation (see Eq.~\eqref{eq:single_impurity_vertex} and \eqref{eq:kmh_lambda_imp_paramagnetic}) reduces Eq.~\eqref{eq:Impurity_vertex_general} to 

\begin{equation}
    \label{eq:impurity_vertex_KMH}
    \begin{aligned}
    &\Lambda_{\text{imp}}^{\sigma\eta}\left(i\omega,i\Omega\right) \\
    &= \frac{\tilde{\chi}_{\text{imp}}^{\sigma\eta}\left(i\omega,i\Omega\right)}{G_{\text{imp}}\left(i\omega+i\Omega\right) G_{\text{imp}}\left(i\omega\right)(1 - \chi^{\eta}_{\text{imp}}\left(i\Omega\right)\mathcal{U}^{\eta}\left(i\Omega\right))}
    \end{aligned}
\end{equation}
and enforces
\begin{align}
    G^{\uparrow}_{\text{imp}} &\overset{!}{=} G^{\downarrow}_{\text{imp}} \equiv G_{\text{imp}}\label{eq:G_imp_paramagnetic}\\
    \sigma_{\sigma\uparrow}^{\eta}\tilde{\chi}_{\text{imp}}^{\uparrow\eta} &\overset{!}{=} \sigma_{\sigma\downarrow}^{\eta}\tilde{\chi}_{\text{imp}}^{\downarrow\eta}\label{eq:chi3_paramagnetic}.
\end{align}
Note that Eq.~\eqref{eq:chi3_paramagnetic} directly follows from rewriting the local version of \eqref{eq:chi3_definition} in terms of creation and annihilation-operators only and by requiring, similar to Eq.~\eqref{eq:chi_loc_ch_z}, all terms with flipped spins to be equal. As there is no other spin-dependency in Eq.~\eqref{eq:impurity_vertex_KMH} but $\tilde{\chi}^{\sigma\eta}_{\text{imp}}$, Eq.~\eqref{eq:kmh_lambda_imp_paramagnetic} directly follows.

\section{Channel-contributions to the bosonic Green's functions for the paramagnetic KMH}
\label{sec:channel_contributions_to_bosonic_gfs}

\begin{figure}
	\begin{center}
		\includegraphics[width=\columnwidth]{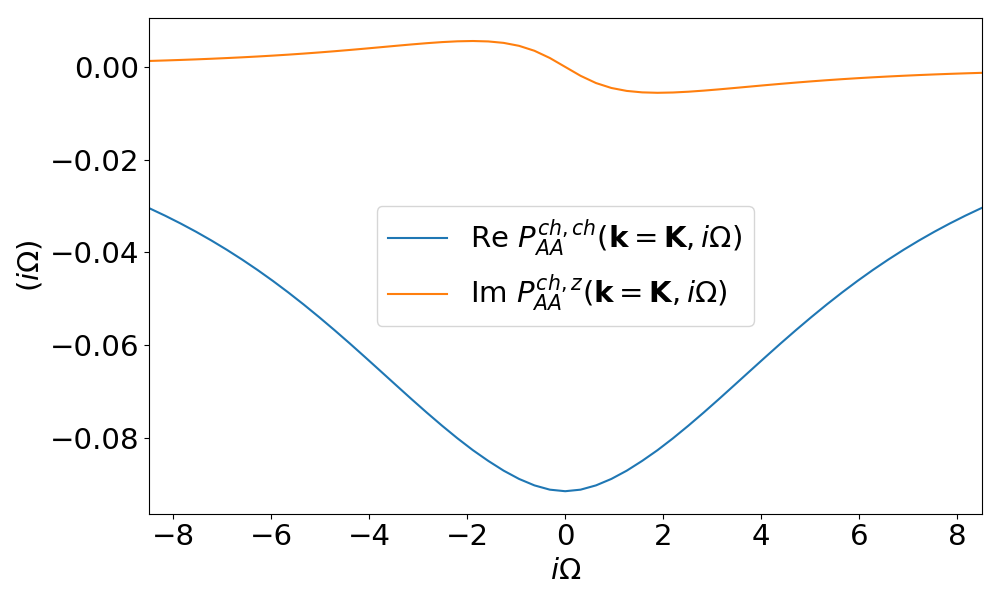}
		\caption{Comparing $P^{ch,ch}$ to $P^{ch,z}$ at the $\mathbf{K}$-Point for $U=7t$ and $\alpha=0.33$. All other contributions, such as $P_{AB}$, Re$P^{z,ch}$ etc. vanish.}
		\label{fig:kmh_polarization_off_diag}
	\end{center}
\end{figure}

\begin{figure}
	\begin{center}
		\includegraphics[width=\columnwidth]{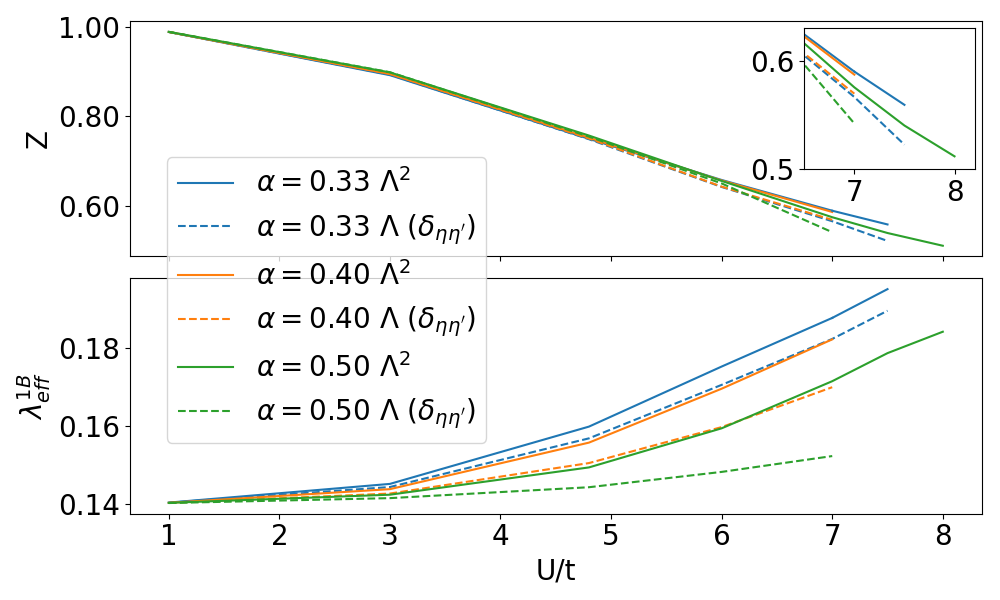}
		\caption{Effective spin-orbit coupling $\lambda^{1B}_{eff}$ for different parameters $U$ and $\alpha$. Comparing also TRILEX $\Lambda^2$ (denoted as $\Lambda^2$) and TRILEX  without channel off-diagonal polarization (denoted as $\Lambda$ ($\delta_{\eta\eta'}$)).}
		\label{fig:kmh_z_and_lambda_eff_alpha_dependent}
	\end{center}
\end{figure}

In order to make a statement about non-vanishing terms of the polarization, we will have a look at the susceptibility first.
It is given by
\begin{equation}
\label{eq:susceptibility}
\begin{aligned}
\chi^{\eta\eta'}_{ij}(\tau) &\equiv \left<T_{\tau}\big(n_i^{\eta} - \big<n_i^{\eta}\big>\big)(\tau)\big(n_j^{\eta'} - \big<n_j^{\eta'}\big>\big)\right>\\
& = \left<T_{\tau}n_i^{\eta}(\tau)n_j^{\eta'}\right> - \big<n_i^{\eta}\big>\big<n_j^{\eta'} \big>,
\end{aligned}
\end{equation}
with space indices $i$ and $j$. In the following we omit the time-ordering Operator $T_{\tau}$ for simplicity. Using the fact that there are \textit{no spin-flip terms} in the Hamiltonian \eqref{eq:kmh_Hamiltonian}) and
\begin{equation}
\begin{aligned}
n^x &\equiv S^{+} + S^{-}\\
n^{y} &\equiv i(S^{-} - S^{+})\\
n^z &\equiv n^{\uparrow} - n^{\downarrow}\\
n^{ch} &\equiv n^{\uparrow} + n^{\downarrow}
\end{aligned}
\end{equation}
leads to
\begin{subequations}
\begin{align}
    &\left<n_i^{x}\right> = \left<n_i^{y}\right> = 0 \\
 	&\chi^{x,ch}_{ij} = \chi^{x,z}_{ij} = \left<(S^{+}_i + S^{-}_i)(\tau)(n^{\uparrow}_j \pm n^{\downarrow}_j)\right> =  0 \label{eq:block_diagonal_suscep1}\\
 	&\rightarrow \chi^{y,ch}_{ij} = \chi^{y,z}_{ij} = 0 \label{eq:block_diagonal_suscep2}\\
 	&\chi^{x,x}_{ij}(\tau) = \chi^{y,y}_{ij}(\tau) = \left<S^{+}_i(\tau)S^{-}_j\right> + \left<S^{-}_i(\tau)S^{+}_j\right>. \label{eq:block_diagonal_suscep3}
\end{align}
 \end{subequations}
  We recognise that the two terms in \eqref{eq:block_diagonal_suscep3} are connected via \textit{particle-hole transformation} as 
 \begin{subequations}
 \begin{align}
    S^{+}_i \rightarrow  -S^{-}_i, \,\,\,\,\,\, S^{-}_i \rightarrow  -S^{+}_i
 \end{align}
 and hence using the \textit{particle-hole symmetry} of the KMH model at \textit{half-filling}
 \begin{align}
 	\chi^{x,y}_{ij}(\tau) = i\left(\left<S^{+}_i(\tau)S^{-}_j\right> - \left<S^{-}_i(\tau)S^{+}_j\right>\right) \overset{p.h.}{=} 0.   
 \end{align}
 \end{subequations}
\textit{Paramagnetic}, on the other hand, means that
\begin{subequations}
\begin{align}
    &\left<n_i^{z}\right> = 0,\\
 	&\left<n_i^{\sigma}(\tau)n_i^{\sigma'}\right> = \left<n_i^{\bar{\sigma}}(\tau)n_i^{\bar{\sigma'}}\right> \rightarrow \chi^{ch,z}_{ii} = \chi^{z,ch}_{ii} = 0, \label{eq:chi_loc_ch_z}\\
 	&\left<S^{+}_i(\tau)S^{-}_i\right> =  \left<S^{-}_i(\tau)S^{+}_i\right> \rightarrow \chi^{x,y}_{ii} = \chi^{y,x}_{ii} = 0.
\end{align}
 \end{subequations}
As susceptibility $\chi$ and polarization $P$ are connected via
\begin{equation}
    \label{eq:connection_susceptibiliy_interaction}
    P^{\eta\eta'}_{\mathbf{k},i\Omega} = - \chi^{\eta\tilde{\eta}}_{\mathbf{k},i\Omega}\left[\delta_{\kappa\kappa'} - U^{\kappa}\chi^{\kappa\kappa'}_{\mathbf{k},i\Omega}\right]^{-1}_{\tilde{\eta}\eta'},
\end{equation}
the block form of $\chi^{\eta\eta'}$ directly transfers to $P^{\eta\eta'}$ and hence also to all other bonsonic Green's functions, justifying Eq.~\eqref{eq:kmh_p_imp_paramagnetic}. \\\\
For the KMH model we find that the biggest contributions to the only non-zero channel-off-diagonal component lie at the $\mathbf{K}(\mathbf{K'})$-point. In order to put this into perspective, in Fig.~\ref{fig:kmh_polarization_off_diag} we compare $P^{z,ch}$ against $P^{ch,ch}$, both at the $\mathbf{K}$-point. We find a ratio between the maxima of both curves of roughly $6\%$.

\begin{figure}
	\begin{center}
		\includegraphics[width=\columnwidth]{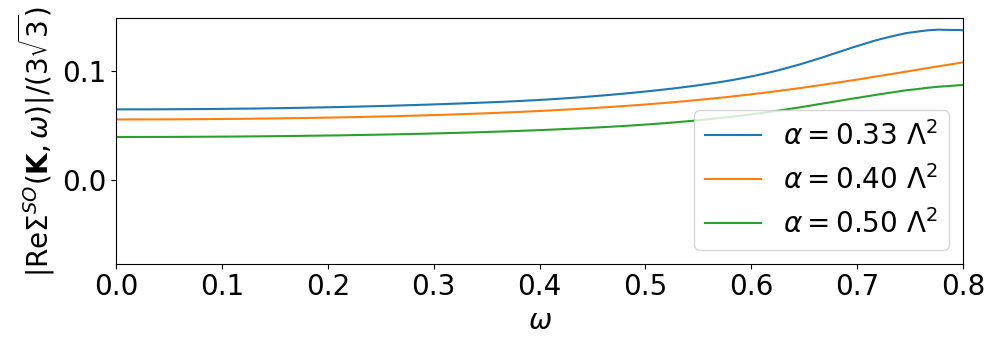}
		\caption{Frequency dependence of SOC-enhancement of the 1-band model (lower figure) for $U =7.0t$ (like Fig.~\ref{fig:kmh_bis_soc_eff_real_freq}) for different $\alpha$. }
		\label{fig:kmh_soc_eff_real_freq_alpha_dependent}
	\end{center}
\end{figure}

\section{Estimations of the SOC-enhancement by analytic approximations}
\label{sec:simpler_approximations}

In the following we apply two analytic approaches, namely Hartree-Fock and second-order perturbation theroy.

\subsection{Hartree-Fock}

For a two-band model the Hartree-Fock decoupling gives a contribution to the effective SOC as \citep{liu_coulomb-enhanced_soc_2008}

\begin{equation}
\label{eq:hartree_2B_model}
	\lambda_\text{eff}^{2B,HF} = \lambda_{SO}^{2B} + \frac{1}{2} \left(U-J\right) \left(\big<n^{\uparrow}_{+}\big> - \big<n^{\uparrow}_{-}\big>\right),
\end{equation}
where $n^\sigma_{\pm}$ are density operators in the eigenbasis (Eq. \eqref{eq:bis_eigenbasis_H_SO}) of the spin-orbit coupling Hamiltonian $H_{SO}$. This contribution is derived from the full interaction Hamiltonian in Hartree-Fock approximation given by

\begin{equation}
\label{eq:hartree_fock_interaction}
	H_{int}^{2B,HF} = \sum_\sigma\left[(U-2J)d^\sigma + (2U-3J)d^{\bar{\sigma}}\right](n^{\bar{\sigma}}_+ + n_-^\sigma),
\end{equation}
where we have defined $d^{\sigma} = \left<n_+^\sigma\right> = \left<n_-^{\bar{\sigma}}\right>$ in order to enforce paramagnetism. 

The enhancement given by Eq. \eqref{eq:hartree_2B_model} with densities extracted from the DMFT solution, must however be equal to the high-frequency ($i\omega \rightarrow \infty$) part of the related self-energy components. Hence, comparing Eq. \eqref{eq:hartree_2B_model} to \eqref{eq:bis_lambda_eff}, the relation
\begin{equation}
    \label{eq:lambda_enhancement_high_frequency}
    \frac{\left(U-J\right)}{2}  \left(\big<n^{\uparrow}_{+}\big> - \big<n^{\uparrow}_{-}\big>\right) = \text{Re}\left(\Sigma^{\uparrow}_+(i\omega_\infty) - \Sigma^{\uparrow}_-(i\omega_\infty)\right)
\end{equation}
must hold, which we can confirm for all calculations presented here.

\begin{figure}
	\begin{center}
		\includegraphics[width=\columnwidth]{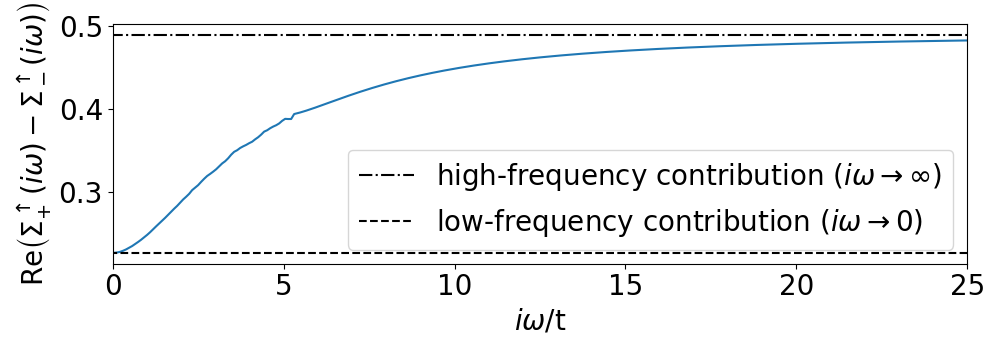}
		\caption{Frequency dependence of the SOC-enhancement for the two-band model for $U=6.7t$ and $J=1.3t$ on the imaginary axis. The high-frequency contribution is calculated from Eq. \eqref{eq:lambda_enhancement_high_frequency}.}
		\label{fig:bis_U6.7t_J1.3t}
	\end{center}
\end{figure}

In Fig. \ref{fig:bis_z_and_lambda_eff} we compare the self-consistent Hartree-Fock solution to the high- and the low-frequency contribution of DMFT and find that Hartree-Fock highly overestimates the enhancement. This is consistent with the fact that there is already a big difference between the high- and the low-frequency contribution of DMFT. This difference is illustrated in Fig. \ref{fig:bis_U6.7t_J1.3t}. Note that we have plotted the frequency-dependence of the enhancement on the imaginary axis, as the analytic continuation becomes increasingly complicated with higher frequencies. Nevertheless, the values at the two points $i\omega\rightarrow 0$ and $i\omega\rightarrow \infty$ are the same on real and imaginary axis.

It is easily seen by looking at Eq. \eqref{eq:kmh_Hamiltonian}, that the Hartree-Fock approximation cannot give a contribution to the spin-orbit coupling of the one-band model, as the interaction and the SOC term do not involve the same bonds. All we get is the term $U\left<n_{i\sigma}\right>$. For a paramagnetic calculation this gives the same constant for every spin (and site) and hence gives only a contribution to the chemical potential. Therefore the Hartree-Fock estimate of the effective SOC is given by the bare SOC:
\begin{equation}
	\label{eq:hartree_1B_model}
	\lambda_\text{eff}^{1B,HF} = \lambda_{SO}^{1B}
\end{equation}
This is in agreement with the high frequency behavior of our TRILEX solution. 

\subsection{Second order perturbation theroy}

\begin{figure}
	\begin{center}
		\includegraphics[width=0.50\columnwidth]{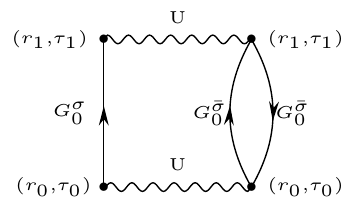}
		\caption{$2^{nd}$ order expansion for single-site model with onsite-interaction only. $G^{\sigma}_0$ is the non-interacting Green's function and $U$ the bare interaction.}
		\label{fig:feynman_2nd_order_bubble}
	\end{center}
\end{figure}

In order to get a non-vanishing contribution to the SOC also in the one band model, we need to include further contributions to the self-energy. Therefore we have a look at the second order contribution to the self-energy. For a single orbital model with only onsite-interaction like the KMH-model this is given by
\begin{equation}
\label{eq:Self_energy_second_order_1B_simply}
\Sigma^{\sigma,2nd}(\mathbf{r},\tau) = -U^{2}G^{\sigma}_0(\mathbf{r},\tau) G^{\bar{\sigma}}_0(\mathbf{r},\tau)G^{\bar{\sigma}}_0(-\mathbf{r},-\tau).
\end{equation}
where $G^{\sigma}_0$ is the non-interacting propagator. The according diagrams are shown in Fig. \ref{fig:feynman_2nd_order_bubble}.
Calculating the contribution to the SOC again as described in sec. \ref{subsec:one_band_model_and_non_local_soc}, for $\lambda_{SO}^{1B} = 0.14t$ we find
\begin{equation}
	\label{eq:effective_SOC_1B_2nd}
	\lambda_\text{eff}^{1B,2nd} = 0.14t + 6.8\cdot 10^{-5}U^{2}t,
\end{equation}
which leads an enhancement of the SOC-coupling that is roughly one order of magnitude smaller than we observe by performing TRILEX calculations. However also this study qualitatively shows that the effective SOC increases when correlations are included.

\section{Some more details on the self-energy}
\label{sec:some_more_details_on_the_self_energy}

\subsection{Fierz-ambiguity of $\lambda_{\text{eff}}^{1B}$}
\label{subsec:alpha-dependence_of_lambda_eff_1B}

As mentioned in subsection \ref{subsec:fixing_the_alpha-parameter}, the choice of the decoupling parameter $\alpha$ can influence the results, especially when single-site impurity problems are solved. In Fig.~\ref{fig:kmh_z_and_lambda_eff_alpha_dependent} we show the development of $\lambda_{\text{eff}}^{1B}$ and $Z$ with increasing local interaction $U$ for different decoupling parameters $\alpha$. We find that while the renormalization factor is affected only weakly by the choice of $\alpha$, $\lambda_{\text{eff}}^{1B}$ decreases in a monotonic fashion when we increase the Fierz parameter from a pure spin decoupling ($\alpha = 1/3$) to a pure charge decoupling ($\alpha = 2/3$). The calculation for $\alpha = 2/3$ turned out to be unstable and, therefore, is not contained in our results, but the tendency becomes clear already for values of $\alpha$ below $2/3$. The qualitative results of the main text remain unchanged, as all calculations for $\alpha > 1/3$ lead to lower effective SOC (see also Fig.~\ref{fig:kmh_bis_lambda_eff_compared}).
We can also see from Fig.~\ref{fig:kmh_z_and_lambda_eff_alpha_dependent} how the dependence of the results on the Fierz-parameter is smaller for TRILEX $\Lambda^2$ than for TRILEX. Mind, however, that we are also ignoring channel-off-diagonal components in the TRILEX approach (see Eq.~\eqref{eq:restriction_TRILEX}) that are small but not vanishing (see Fig.~\ref{fig:kmh_polarization_off_diag}).

We note in passing that it has been argued that the Fierz-ambiguity can be resolved within the dual TRILEX approach~\cite{double_Lambda2}.

\subsection{Frequency-dependence of SOC-enhancement}

For the sake of completeness we also look at the frequency dependence of the SOC-enhancement for different Fierz-parametersi.e., we did analytic continuations of $\textrm{Re}\,\Sigma^{SO}$ as we did in the main text, but for different $\alpha$ parameters, which is shown in  Fig.~\ref{fig:kmh_soc_eff_real_freq_alpha_dependent}. We see clearly that all calculations for different $\alpha$ follow a very similar trend, namely a constant value for small frequencies followed by a small upturn.

\subsection{Effect of SOC on the normal non-SOC part  of the self-energy}

An interesting question is how the inclusion of SOC changes the normal, i.e. time-reversal symmetry preserving part of the self-energy.  
In the two-band model, this symmetry preserving part is given by $\frac{1}{2}\left( \Sigma_{+}^{\uparrow} + \Sigma_-^{\uparrow}\right)$. For the one-band model, both $\Sigma^d$ and $\Sigma_{AB}$ are the normal components. In order to make a comparison between the two models we will, however, solely focus on the impurity self energies, which for the one-band case is $\Sigma_{\text{imp}} = \sum_{\mathbf{k}}\Sigma^d$.
\begin{figure}
	\begin{center}
		\includegraphics[width=\columnwidth]{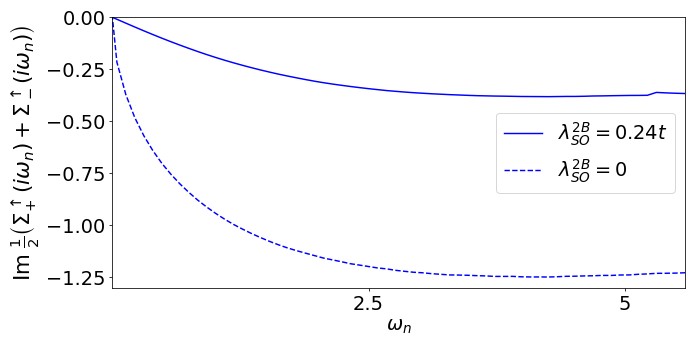}
		\includegraphics[width=\columnwidth]{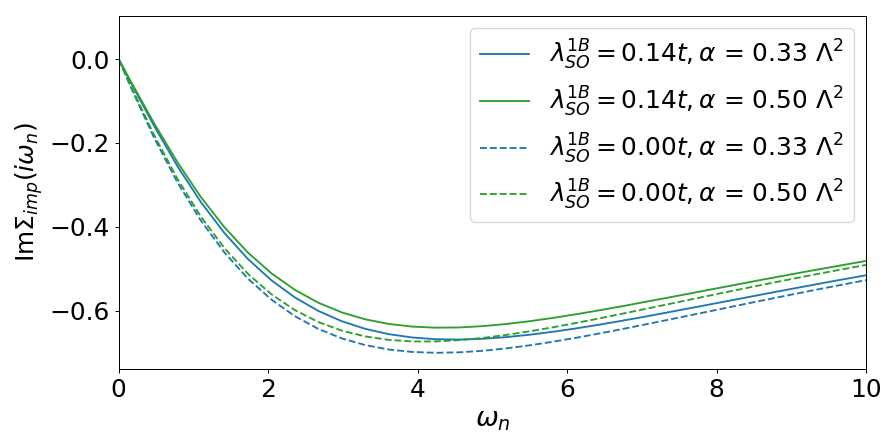}
		\caption{Im $\frac{1}{2}\left( \Sigma_{+}^{\uparrow} + \Sigma_-^{\uparrow}\right)$ of the 2-band model at $U=12t$ and $J=2t$, with and without SOC (upper panel). $\Sigma_{\text{imp}}$ of the 1-band model at $U=4.8t$, for different $\alpha$-values with and without SOC (lower panel).}
		\label{fig:kmh_bis_Im_sigma_with_and_without_soc}
	\end{center}
\end{figure}

Fig.~\ref{fig:kmh_bis_Im_sigma_with_and_without_soc} shows the effect of introducing a SOC-term in both models. We can see that in either case the correlations are reduced when $\lambda_{SO}$ is turned on. However, for the two-band model for bismuthene this effect is much stronger than for the KMH model. This goes hand in hand with a much larger impact of correlations on the effective SOC in the two-band model, as discussed in the main text. Obviously, here the inversion of the arguments is true, namely that not only correlations have a strong impact on the effective SOC, but also that SOC has a strong impact on correlations related to normal components of the self-energy.
We notice in passing that the choice of the decoupling parameter $\alpha$ has a rather small quantitative impact on this qualitative result.

\begin{figure}
	\begin{center}
		\includegraphics[width=.95\columnwidth]{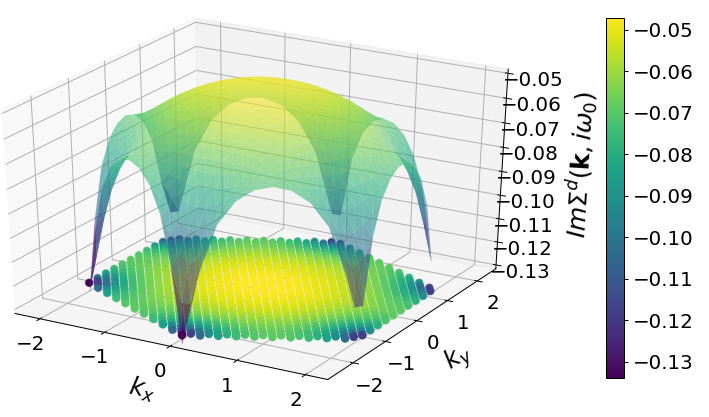}
		\includegraphics[width=.95\columnwidth]{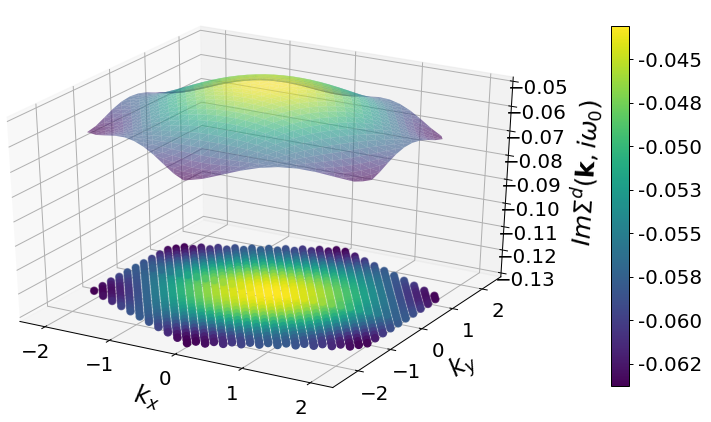}
		\caption{$\text{Im}\, \Sigma^{d}(\mathbf{k},i\omega_0)$ at $U=4.8t$ and $\alpha =0.33$ for $\lambda^{1B}_{SO} = 0$ (upper figure) and $\lambda^{1B}_{SO} = 0.14t$ (lower figure).}
		\label{fig:kmh_Im_sigma_diag_in_k_space}
	\end{center}
\end{figure}

\subsubsection{$\mathbf{k}$-dependence of $\Sigma^{d}$}

Last but not least we want to look at the effect of $\lambda_{SO}^{1B}$ on the $\mathbf{k}$-dependence of $\Sigma^d$. From Fig.~\ref{fig:kmh_Im_sigma_diag_in_k_space} we find that SOC reduces the $k$-dependence of $\textrm{Im}\,\Sigma^d$ and hence again the strength of the correlation.
\begin{figure}
	\begin{center}
		\includegraphics[width=\columnwidth]{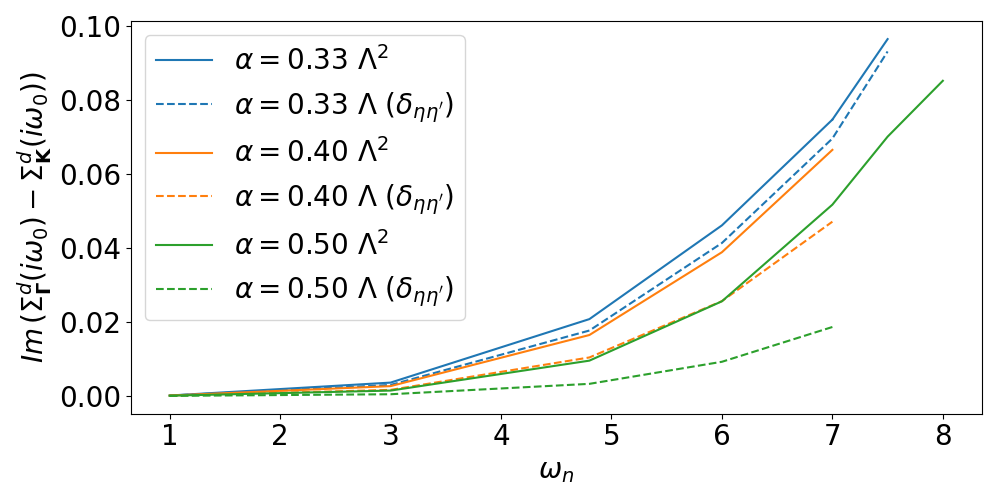}
		\caption{Difference of the two most distant points in $\text{Im}\,\Sigma^d(\mathbf{k},i\omega_0)$ $\Gamma$ and $K$ as a measure of the $\mathbf{k}$-dependence of $\Sigma^d$ for $\lambda_{SO}^{1B}$ and different parameters of $U$ and $\alpha$. We again compare TRILEX $\Lambda^2$ to TRILEX as in Fig.~\ref{fig:kmh_z_and_lambda_eff_alpha_dependent}.}
		\label{fig:kmh_k_dependency_alpha_dependent}
	\end{center}
\end{figure}
In order to show the development of the $\mathbf{k}$-dependence with increasing interactions we introduce $\text{Im}\, (\Sigma^{d}_{\mathbf{\Gamma}}(i\omega_n) - \Sigma^{d}_{\mathbf{K}}(i\omega_n))$ as a measure for the non-locality of the self-energy. We consider this reasonable as it involves the two most different values in the Brillouin zone. Fig.~\ref{fig:kmh_k_dependency_alpha_dependent} shows this measure for different interaction values $U$ and decouplings $\alpha$. Not surprisingly, the $\mathbf{k}$-dependence increases with increasing $U$, as correlations become stronger.
Contrary to Fig.~\ref{fig:kmh_bis_Im_sigma_with_and_without_soc}, but similar to Fig.~\ref{fig:kmh_z_and_lambda_eff_alpha_dependent}, we find that the relative effect of $\alpha$ on the $\mathbf{k}$-dependence is clearly visible in this measure. This again shows that for the KMH model non-local quantities are more affected by the choice for the decoupling parameter than local quantities. However, we again find that the treatment in TRILEX $\Lambda^2$ decreases the Fierz-ambiguity as compared to TRILEX.

\bibliography{honeycomb}

\end{document}